\documentclass[journal]{IEEEtran}

\usepackage[bookmarks,colorlinks]{hyperref}
\usepackage[linesnumbered,ruled,lined]{algorithm2e}
\usepackage{enumitem}
\usepackage{algpseudocode}
\usepackage{cite}
\usepackage{amsmath,amssymb,amsfonts,amsthm,steinmetz}
\usepackage{graphicx}
\usepackage{mathrsfs}  
\usepackage{textcomp}
\usepackage{acronym}
\usepackage{xcolor}
\usepackage{upgreek,xspace}

\usepackage{esvect}

\usepackage{times}
\usepackage{bm}
\usepackage{amsmath}
\usepackage{amssymb}
\usepackage{stmaryrd}
\usepackage{babel}
\usepackage{graphics, graphicx}
\usepackage{xcolor}
\usepackage{gensymb}
\usepackage{cite}
\usepackage{enumitem}
\usepackage{url}
\newtheorem{rem}{Remark}

\usepackage{nicematrix}
\usepackage{comment}
\usepackage{soul}


\newcommand{\mI}{\ensuremath{\mathbf{I}}}
\newcommand{\mO}{\ensuremath{\textcolor{gray}{\mathbf{0}}}}
\newcommand{\iC}{\ensuremath{\mathcal{C}}}
\newcommand{\iN}{\ensuremath{\mathcal{N}}}
\newcommand{\iP}{\ensuremath{\mathcal{P}}}

\newcommand{\mS}{\ensuremath{\mathbf{S}}}
\newcommand{\mZ}{\ensuremath{\mathbf{Z}}}

\newcommand{\X}{\ensuremath{\mathrm{X}}}
\newcommand{\Y}{\ensuremath{\mathrm{Y}}}
\newcommand{\super}{\ensuremath{\mathrm{sup}}}
\newcommand{\con}{\ensuremath{\mathrm{con}}}
\newcommand{\G}{\ensuremath{\mathrm{G}}}
\newcommand{\U}{\ensuremath{\mathrm{U}}}
\newcommand{\V}{\ensuremath{\mathrm{V}}}
\newcommand{\XUV}{\ensuremath{\mathbf{X}^\mathrm{UV}}}
\newcommand{\XVU}{\ensuremath{\mathbf{X}^\mathrm{VU}}}
\newcommand{\A}{\ensuremath{\mathrm{A}}}
\newcommand{\B}{\ensuremath{\mathrm{B}}}
\newcommand{\C}{\ensuremath{\mathrm{C}}}
\newcommand{\D}{\ensuremath{\mathrm{D}}}
\newcommand{\meta}{\ensuremath{\mathrm{M}}}

\newcommand{\metaref}{\ensuremath{\mathrm{ABCD}_\mathrm{ref}}}
\newcommand{\iI}{\ensuremath{\mathcal{I}}}
\newcommand{\iJ}{\ensuremath{\mathcal{J}}}

\begin{document}

\title{Updatable Closed-Form Evaluation of Arbitrarily Complex Multi-Port Network Connections}

\author{Hugo~Prod'homme~and~Philipp~del~Hougne,~\IEEEmembership{Member,~IEEE}
\thanks{
H.~Prod'homme and P.~del~Hougne are with Univ Rennes, CNRS, IETR - UMR 6164, F-35000, Rennes, France (e-mail: \{hugo.prodhomme; philipp.del-hougne\}@univ-rennes.fr).
}
\thanks{
P.d.H. acknowledges funding from CREACH LABS (project AdverPhy), the ANR France 2030 program (project ANR-22-PEFT-0005), and the ANR PRCI program (project ANR-22-CE93-0010).
}
\thanks{\textit{(Corresponding Author: Philipp del Hougne.)}}
}

\maketitle

\begin{abstract}
The design of large complex wave systems (filters, networks, vacuum-electronic devices, metamaterials, smart radio environments, etc.) requires repeated evaluations of the scattering parameters resulting from complex connections between constituent subsystems. Instead of starting each new evaluation from scratch, we propose a computationally efficient method that updates the outcomes of previous evaluations using the Woodbury matrix identity. To enable this method, we begin by identifying a closed-form approach capable of evaluating arbitrarily complex connection schemes of multi-port networks. We pedagogically present unified equivalence principles for interpretations of system connections, as well as techniques to reduce the computational burden of the closed-form approach using these equivalence principles. Along the way, we also achieve the closed-form retrieval of the power waves traveling through connected ports. We illustrate our techniques considering a complex meta-network involving serial, parallel and cyclic connections between multi-port subsystems. We further validate all results with physics-compliant calculations considering graph-based subsystems, and we conduct exhaustive statistical analyses of computational benefits originating from the reducibility and updatability enabled by our approach. Finally, we find that working with scattering parameters (as opposed to impedance or admittance parameters) presents a fundamental advantage regarding an important class of connection schemes whose closed-form analysis requires the treatment of some connections as delayless, lossless, reflectionless and reciprocal two-port scattering systems. We expect our results to benefit the design (and characterization) of large composite (reconfigurable) wave systems.
\end{abstract}


\section{Introduction}\label{sec:Introduction}

The need to evaluate the network parameters of connections between multi-port scattering systems arises in diverse contexts. A prime example of this need is found in the design of large wave systems. Simulating the whole system would be prohibitively costly, but the finite number of components of which the system is composed can be individually simulated at a reasonable cost. Once each individual component's network parameters are obtained, the network parameters of a complex connection of these individual components is thus sought. This approach, known as ``diakoptics''~\cite{happ1974diakoptics,klos1982diakoptics}, is taken in areas spanning from the design of filters, networks and vacuum-electronic devices~\cite{shirakawa1968synthesis,koziel2009accelerated,cheng2009space,cheng2010progress,Snyder_NEMO,jabotinski2018efficient,snyder2021emerging} to the design of metamaterials~\cite{TAP_tensor_grbic,TAP_DomainDecomposition}.

Another example of contemporary interest are parametrized wireless communications channels found in ``smart radio environments'' equipped with reconfigurable intelligent surfaces (RISs). An RIS is an array of elements with tunable scattering properties; typically, an RIS is implemented as an array of patch resonators equipped with individually tunable lumped elements such as PIN diodes. A physics-compliant representation of the smart radio environment treats the radio environment as a multi-port system comprising the antenna ports as well as auxiliary ports at the locations of the RIS's tunable lumped elements, and these auxiliary ports are then terminated by tunable individual loads~\cite{tapie2023systematic}.\footnote{The recent literature contains many works~\cite{gradoni_EndtoEnd_2020,shen2021modeling,faqiri2022physfad,tap2022,badheka2023accurate,sol2023experimentally,mursia2023,tapie2023systematic,nossek2024,rabaultTWC,nerini2024universal,abrardo2024} on variants of the general formulation given in Ref.~\cite{tapie2023systematic}. Many of these variants make assumptions about the radio environment (e.g., that it is free space or only composed of point-like scatterers), the structural scattering of antennas and/or RIS elements (e.g., that it is negligible) and/or mathematical simplifications (e.g., the ``unilateral approximation''~\cite{ivrlavc2010toward} which assumes that certain off-diagonal blocks of the impedance matrix are zero); the diverse variants also differ regarding their choice of equivalent representations in terms of scattering parameters, impedance parameters or a coupled-dipole formalism.} This representation hence involves a connection between two multi-port systems, one of which has a diagonal scattering matrix. Incidentally, the same concept also underpins the so-called ``port tuning'' method during the design closure of filter synthesis~\cite{koziel2009accelerated,cheng2009space,cheng2010progress}, as recently pointed out in Ref.~\cite{faul2024agile}. The extension to ``beyond-diagonal'' RIS (BD-RIS) proposed in Ref.~\cite{shen2021modeling} can be understood as terminating the auxiliary ports of the radio environment with a tunable load circuit that is itself a multi-port network terminated by tunable individual loads~\cite{del2024physics}, resulting in a chain cascade of three multi-port networks, the last of which has a diagonal scattering matrix.
The extension to ``stacked intelligent metasurfaces'' (SIM) proposed in Ref.~\cite{SIM_intro} also amounts to a chain cascade of multi-port networks in a physics-compliant representation~\cite{SIM_clerckx}.

A red line running through all of the above examples is the need to repeatedly evaluate the multi-port network representation of a fixed connection scheme of multi-port networks. Indeed, all examples are concerned with the optimization of some components within a connection of multi-port networks such that the  transfer function resulting from the connection approximates desired properties. Irrespective of the details of the chosen optimization procedure, many evaluations of the considered connection of multi-port networks are generally required, and they likely differ only regarding a few details of the involved components. Yet, each evaluation may be computationally very costly because it generally involves matrix inversions (see below). Instead of performing each evaluation from scratch, it would be highly desirable to merely update previous evaluations of the considered multi-port network connection. 

A prerequisite for setting up a universal technique for updating arbitrarily complex connections of multi-port networks is a transparent closed-form method for evaluating them in the first place. Because such a method is to date not widely known, the first part of this paper is dedicated to illustrating a transparent closed-form evaluation method prior to studying its updatability. 

Of course, the simplest case of cascading two two-port networks by connecting one port from each is covered in standard textbooks~\cite{Pozar2011,dobrowolski2010microwave} based on the ABCD representation of the two multi-port networks; this approach is implemented in Matlab's RF Toolbox. Indeed, after converting the network parameters to ABCD matrices, these matrices can simply be multiplied and converted back to scattering or impedance parameters. However, the applicability of this ABCD-based method is limited to chain cascades of systems with the same number of unconnected and connected ports such that half of the ports of the first system are connected to half of the ports of the second system. More complex connection schemes involving different numbers of unconnected and connected ports and/or featuring more than two sets of ports would require an iterative application of the ABCD chain cascade with numerous conversions back and forth between scattering or impedance and ABCD parameters. Such an approach is hence intransparent and not amenable to a closed-form description, and its application is generally prone to implementation errors. 

In fact, the evaluation of a connection of multi-port systems does not require any conversion between different network representations (scattering parameters, impedance parameters, ABCD parameters, etc.). A number of alternative approaches exists. Signal flow graphs~\cite{hunton1960SignalFlowGraphs} are one such alternative but it is difficult to apply them to large complex problems. In addition, various iterative approaches exist. On the one hand, one can iteratively evaluate complex connections as sequential cascades of two systems~\cite{anderson_cascade_1966,filipsson_new_1981,simpson_generalized_1981,belenguer_krylovs_2013,caballero2014extending}. On the other hand, one can define a supersystem comprising all involved subsystems (such that the supersystem is described by a block-diagonal scattering matrix) and then evaluate one-by-one the inner connections of this supersystem~\cite{compton_perspectives_1989, filipsson_new_1981} (see Sec.~\ref{subsec_terminology} for a rigorous definition of inner connection); this approach is utilized in the current Python-based scikit-rf library. These iterative approaches do not provide a closed-form solution that would be amenable to updatability. Finally, there are some single-step algorithms~\cite{monaco1974,chu_generalized_1986,yang_new_2018,de2022single} that are typically studied in limited contexts like chain cascades rather than for more complex connection schemes (such as the meta-network described in Fig.~\ref{fig:meta-network}). Overall, a unifying and transparent approach to evaluate arbitrarily complex connection schemes for multi-port networks in closed-form is not widely known to date. We illustrate such an approach in a pedagogical manner and with detailed analysis in this paper.

Then, provided that a closed-form approach to evaluating the connection of the multi-port systems exists, the Woodbury matrix identity~\cite{hager1989updating} will provide a computationally efficient and accurate manner of updating previous evaluations. To date, however, this efficient updating method has only been studied for the most basic cascade problem encountered for RIS-parametrized wireless channels~\cite{prod2023efficient}.\footnote{Related applications of the Woodbury identity in electromagnetic contexts not explicitly concerned with multi-port system connections can be found in Refs.~\cite{boag1996design,chen2016accelerated,chen2018analysis,budhu2023fast}.} To the best of our knowledge, no comparable technique for arbitrarily complex connections between multi-port systems has been studied to date, presumably, due to a lack of transparent closed-form techniques to evaluate these connections. The present paper is dedicated to establishing an \textit{updatable closed-form} technique to evaluate \textit{arbitrarily complex} multi-port network connections.

The application scope of the updating technique that we study in this paper may extend well beyond our current motivation originating from considerations of computational efficiency. On the one hand, the infinitesimal limit of updates are derivatives which play an important role in optimization methods (e.g., gradient-descent methods such as backpropagation)~\cite{director1969automated,director1969generalized,iuculano1971network,monaco1974,nikolova2004adjoint}. On the other hand, we expect our updating scheme to enable the derivation of closed-form parameter estimation techniques for tunable composite wave systems. Indeed, for the simpler special case of RIS-parametrized wireless channels, updating schemes have enabled closed-form physics-compliant end-to-end channel estimation, as well as an extension to the conceptually closely related ``virtual vector network analyzer'' that estimates a multi-port scattering matrix by terminating a (large) subset of the ports with tunable loads~\cite{del2024minimal,del2024virtual}. Another related example is pilot-free phase-conjugation focusing on a moving user equipment~\cite{sol2024optimal}.

\subsection{Contributions and Organization}

In the remainder of this paper, one section is dedicated to each of our main contributions:\footnote{All sections except for Sec.~\ref{sec_impedance_admittance} use scattering parameters.} 
\begin{enumerate}
    \item \textit{Equivalence principles for system connections.} In Sec.~\ref{sec:Equivalencies}, we highlight and illustrate multiple equivalent ways of interpreting connections between multi-port systems. These include the equivalence between ``inner'' and ``outer'' connections, as well as a representation in terms of inner connections of a supersystem. Parts of these principles appear implicitly in earlier studies, but our unified and synthetic perspective on these principles has an important pedagogical value and sets the scene for the rest of our paper. We also highlight that the equations for cascade loading and the Redheffer star product can be derived from each other.
    \item \textit{Global method for arbitrarily complex connections.} In Sec.~\ref{sec:global}, we illustrate our transparent closed-form global method for the case of a highly complex connection scheme involving parallel junctions, serial chains and cycles. We refer to the resulting system as a ``meta-network''. 
    \item \textit{Reducibility.} In Sec.~\ref{sec:reducibility}, we discuss reducibility principles (based on Sec.~\ref{sec:Equivalencies}) that interpret some of the connected systems as connections to reduce the mathematical complexity. We discuss the extent to which complex connection schemes are reducible and illustrate these conclusions with the meta-network example, including an analysis of possible gains in computational efficiency.
    \item \textit{Updatability.} In Sec.~\ref{sec:updatability}, we study the updatability of  previously evaluated (arbitrarily complex) connection schemes upon a change of one of the constituent systems. This is the most important contribution of the present paper. We illustrate the technique with various examples involving the meta-network, and we evaluate the gains in computational efficiency enabled by the technique. 
    \item  \textit{Closed-form recovery of power waves passing through connected ports.} In Sec.~\ref{sec:generalizedconnect_potentials}, we demonstrate that our approach enables the closed-form evaluation of the power waves travelling through the connected ports, which is important to assess system vulnerabilities and, subject to sufficient a priori knowledge, to reconstruct internal field distributions. This method is compatible with the reducibility and updatability of our approach from Sec.~\ref{sec:reducibility} and Sec.~\ref{sec:updatability}, respectively.
    \item \textit{Transposition to impedance and admittance parameters.} In Sec.~\ref{sec_impedance_admittance}, we transpose the global method from Sec.~\ref{sec:global} from scattering parameters to impedance and admittance parameters. We highlight a difficulty in describing a $\delta$-connection (see Sec.~\ref{subsec_terminology} for a definition) in terms of impedance or admittance parameters, as well as a work-around and its penalty in terms of numerical accuracy. Working with scattering parameters therefore presents a fundamental benefit when working with important classes of connection schemes that are not fully reducible (see detailed definition in Sec.~\ref{sec:reducibility}). For completeness, we further provide equations for cascade loading as well as the closed-form retrieval of voltages at connected ports and currents travelling through connected ports in terms of impedance and admittance parameters.
\end{enumerate}
In Sec.~\ref{sec_conclusion}, we briefly conclude.

\subsection{Validation}
\label{subsec_val}

All results in this paper are validated numerically for multi-port systems that are ideal transmission-line networks (hereafter referred to as ``graphs''). On the one hand, the scattering matrix of such a graph is known analytically. Its computational evaluation is orders of magnitude faster than a finite-element simulation or an experimental measurement. Therefore, we can consider many examples of physics-compliant scattering matrices within a reasonable time. On the other hand, the connection of multiple graphs is itself a graph whose scattering matrix is known analytically~\cite{kostrykin_generalized_2001}. Therefore, we can evaluate the ground truth independently, analytically, and free of any uncertainty.

The formalism used to compute the scattering matrix of a graph is well-documented in the literature. It can be derived in terms of the telegrapher's equations or the one-dimensional Schrödinger equation on a quantum graph; the equivalence between the two approaches has been worked out explicitly in Ref.~\cite{hul_experimental_2004}. Following the quantum-graph formalism~\cite{texier_scattering_2001,kottos_quantum_2003, kuchment_quantum_2005}, we detail in Appendix~\ref{Appendix_graphs} how to determine the scattering matrix of a graph and how to retrieve the power waves traveling on subgraphs.

\subsection{Terminology and definitions}
\label{subsec_terminology}

Throughout this paper, we assume that the considered scattering systems are linear, time-invariant and passive, and that their ports are monomodal\footnote{Multimodal ports can be treated as sets of multiple monomodal ports within our framework~\cite{multimodeTRL}.}. Hence, an $N$-port system X is fully characterized by its scattering matrix $\mathbf{S}^\mathrm{X} \in \mathbb{C}^{N \times N}$, or its impedance matrix $\mathbf{Z}^\mathrm{X} \in \mathbb{C}^{N \times N}$, or its admittance matrix $\mathbf{Y}^\mathrm{X} \in \mathbb{C}^{N \times N}$ (see Appendix~\ref{Appendix_Definitions} for definitions). The superscripts identify the corresponding system (here: X).

We define a \textit{free} port as a port that is not connected to any other port. We refer to a connection between ports as an \textit{inner} connection if all concerned ports belong to the same system, and as an \textit{outer} connection otherwise.
The (inner or outer) connection between $m$ ports can itself be interpreted as an $m \times m$ scattering system characterized by its scattering matrix $\mathbf{S}_\mathrm{con} \in \mathbb{C}^{m \times m}$.

We refer to a delayless, lossless, reflectionless and reciprocal connection between a pair of ports as a \emph{$\delta$-connection}. These combined properties of a $\delta$-connection imply that it has an overall ``neutral'' effect on signal transmission (in some sense akin to the addition of zero or the multiplication with unity in basic algebra), allowing a $\delta$-connection to be added or removed from a connection of systems without altering the resulting behavior. Therefore, any connection can be broken down into a description involving $\delta$-connections between connected ports of the subsystems.

\subsection{Notation}

We use the notation $\mathbf{S}^\mathrm{X}_\mathcal{AB}$ to refer to the block of $\mathbf{S}^\mathrm{X}$ that corresponds to the input ports and output ports whose indices are comprised in the (not necessarily contiguous) sets of port indices $\mathcal{A}$ and $\mathcal{B}$. 

We distinguish between free and connected ports as follows:
 \begin{itemize}
     \item $\iN_\X$ are the free ports of system $\X$.
     \item $\iC_\X$ are the connected ports of system $\X$.
     \item $\iC_\X^\Y$ are the ports of system $\X$ connected to system $\Y$.
     \item $\iP_\X = \iN_\X \cup \iC_\X$ are all ports of system $\X$.
 \end{itemize}
 The writing of partitioned scattering matrices can hence be refined as follows:
\begin{equation*}
\mS^\mathrm{X} = 
\begin{bNiceMatrix}
\mS_{\iN_\X\iN_\X}^\mathrm{X} & \mS_{\iN_\X\iC_\X}^\mathrm{X} \\
\mS_{\iC_\X\iN_\X}^\mathrm{X} & \mS_{\iC_\X\iC_\X}^\mathrm{X} \\
\end{bNiceMatrix} = \,
\begin{bNiceMatrix}[first-row, first-col]
    & \iN_\X               & \iC_\X   \\
\iN_\X & \Block{2-2}{\mS^\mathrm{X}}  &       \\
\iC_\X &                   & 
\end{bNiceMatrix}.
\label{eq_block}
\end{equation*}

If a connection system composed of $\delta$ connections is inserted between the connected ports of systems X and Y (recall that $\delta$ connections between connected system ports can be added or removed at wish given their overall neutral effect), we denote by $\hat{\mathcal{C}}_{\mathrm{X}}^{\mathrm{Y}}$ the ports of this connection system that are connected to the ports $\iC_\X^\Y$ of system X. We indicate such connection systems composed of $\delta$ connections in grey color in our schematics to emphasize their neutral effect.

$\mI$ and $\mathbf{0}$ denote the identity matrix and the null matrix, respectively. When they appear as blocks inside a larger matrix, their dimensions correspond to those of their respective blocks. The blocks containing $\mI$ are always square. The $\mathbf{0}$ matrices are written in grey inside blocks to facilitate reading.

$n(\dots)$ designates the cardinality of a set (e.g., implying $\mS^\X_\mathcal{AB}\in\mathbb{C}^{n(\mathcal{A})\times n(\mathcal{B})}$).
$\varnothing$ refers to a null set.
$\setminus$ is the set difference operator.
$\mathbf{A}^T$ is the transpose of $\mathbf{A}$.
$\mathbf{A}^\dagger$ is the transpose conjugate of $\mathbf{A}$. 
$\mathrm{blockdiag}\left\{\mathbf{B},\mathbf{C},\mathbf{D}\right\}$ defines a blockdiagonal matrix whose diagonal blocks are $\mathbf{B}$, $\mathbf{C}$ and $\mathbf{D}$.
$\otimes$ denotes the Kronecker product.
$\delta_{ij}$ is the Kronecker delta.
$\nabla$ is the gradient operator.
$\jmath$ is the imaginary unit.

\section{Equivalence principles for system connections}
\label{sec:Equivalencies}

In this section, we establish the equivalence between different ways of interpreting connections between multi-port systems. 

\subsection{Cascade loading}

We start by recalling the well-known~\cite{anderson_cascade_1966, ha1981solid, ferrero1992new, de2010signal, Hochwald2014, yang_new_2018, li2024beyond, reveyrand2018multiport, del2024minimal, del2024virtual, Viikari2024} ``cascade loading'' formula for scattering parameters; a derivation is provided in Appendix~\ref{Appendix_Derivations_S} for completeness.
We consider a system $(1)$ whose ports are collected in the set $\iP=\iN\cup\iC$; the ports $\iC$ are connected to the $n\left(\iC\right)$ ports of a system $(2)$. The scattering matrix $\mS^{(12)}$ of the connected system is given by the following ``cascade loading'' formula for scattering parameters:
\begin{equation}
    \mS^{(12)} = \mS^{(1)}_{\iN\iN} + \mS^{(1)}_{\iN\iC} 
    \left({\mS^{(2)}}^{-1} - \mS^{(1)}_{\iC\iC}\right)^{-1} 
    \mS^{(1)}_{\iC\iN}.
\label{eq:matrix_loading}
\end{equation}

\begin{figure}[h]
    \centering
    \includegraphics[width=\columnwidth]{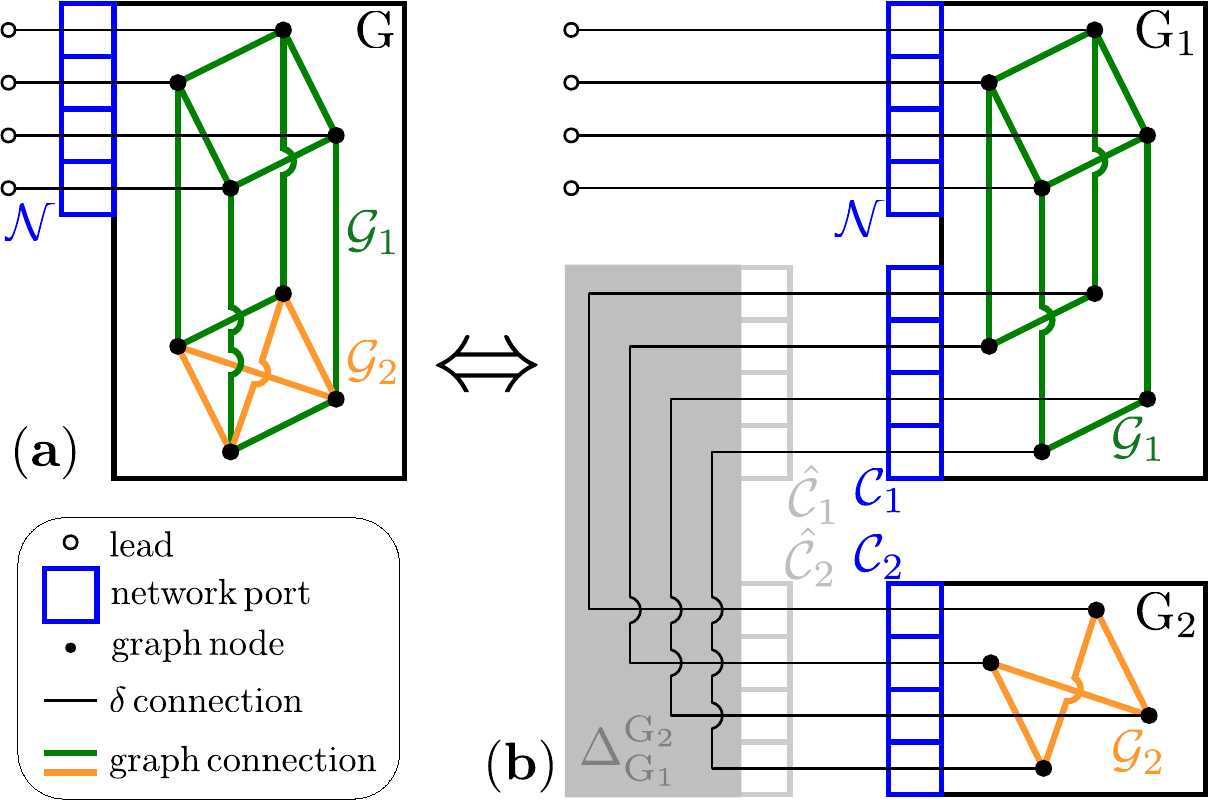}
    \caption{Illustration of the inner-outer equivalency. (a) Representation of a network $\G$ with scattering properties inherited from the underlying graph structure, divided into two interconnected subgraphs $\mathcal{G}_1$ and $\mathcal{G}_2$. (b) Connection of two networks $\G_1$ and $\G_2$, each inheriting its scattering properties from the corresponding underlying subgraph. The $\delta$-connections are delayless, lossless, reflectionless and reciprocal. The network $\G$ and the connection of the networks $\G_1$ and $\G_2$ exhibit the same scattering properties, which are probed via the non-connected free ports $\iN$.}
    \label{fig:innnerouterequivalency}
\end{figure}

\subsection{Inner-outer equivalency}

The role of system (2) in this cascade loading can be interpreted in two equivalent ways. On the one hand, one can interpret (2) as constituting a (potentially complex) set of ``inner'' connections between some of the ports of (1). On the other hand, one can interpret (2) as a system on its own in which case there are ``outer'' connections between distinct pairs of ports, each pair including one port from (1) and one port from (2). 

It is instructive to illustrate this ``inner-outer equivalency'' by considering for concreteness a realization of systems (1) and (2), respectively, as graphs $\mathcal{G}_1$ and $\mathcal{G}_2$ (i.e., networks of transmission lines), in line with the sort of scattering systems we consider for validation in this paper. The scattering systems associated with graphs $\mathcal{G}_1$ and $\mathcal{G}_2$ are denoted by $\mathrm{G}_1$ and $\mathrm{G}_2$, respectively. The scattering matrix of the system $\mathrm{G}$ encasing the gluing of $\mathcal{G}_1$ and $\mathcal{G}_2$ is equal to the cascaded scattering matrices of the systems  $\mathrm{G}_1$ and $\mathrm{G}_2$~\cite{kostrykin_kirchhoffs_1999}.

An illustration of the inner-outer equivalency for such graph-based realizations of (1) and (2) is provided in Fig.~\ref{fig:innnerouterequivalency}. The interpretation of $\mathrm{G}_2$ as ``inner'' connection of $\mathrm{G}_1$ is depicted in Fig.~\ref{fig:innnerouterequivalency}(a); the interpretation of the same setup as an ``outer'' connection between $\mathrm{G}_1$ and $\mathrm{G}_2$ is depicted in Fig.~\ref{fig:innnerouterequivalency}(b). 
By analogy with Eq.~(\ref{eq:matrix_loading}), it follows immediately that
\begin{equation}
    \mS^{\G} = \mS^{\G_1}_{\iN\iN} + \mS^{\G_1}_{\iN\iC_1} \left(\left({\mS^{\G_2}}\right)^{-1} - \mS^{\G_1}_{\iC_1\iC_1}\right)^{-1}  \mS^{\G_1}_{\iC_1\iN}.
\label{eq:innerouter-a}
\end{equation}

\subsection{Scattering system of $\delta$-connections}
\label{sec:connections_as_scattering}

The inner-outer equivalency can be further applied to the study case of Fig.\,\ref{fig:innnerouterequivalency}(b). Indeed, the set of $\delta$-connections between the ports $\iC_1$ and $\iC_2$ can be considered as a set of internal connections within a supersystem $\G_{12}$ comprising $\G_1$ and $\G_2$. Therefore, the set of $\delta$-connections can be considered as a scattering system on its own, denoted $\Delta_{\G_1}^{\G_2}$ (highlighted as a grey overlay in Fig.\,\ref{fig:innnerouterequivalency}(b)); recall that the $\delta$-connections have a ``neutral'' effect, as clarified in Sec.~\ref{subsec_terminology}. The scattering matrix of this system composed of $\delta$-connections\footnote{Although implicitly underlying the analysis presented in several studies~\cite{anderson_cascade_1966,monaco1974,de2022single,ranjbar_analysis_2017}, the scattering matrix of $\delta$-connections has never been expressly described, to the best of our knowledge.}, denoted by $\Delta_{\G_1}^{\G_2}$, is given by
\begin{equation}
\mS^{\Delta_{\G_1}^{\G_2}} = \begin{bNiceMatrix}[first-row, first-col]
            & \hat{\mathcal{C}}_1 & \hat{\mathcal{C}}_2      \\
\hat{\mathcal{C}}_1       & \mO   & \mI        \\
\hat{\mathcal{C}}_2       & \mI   & \mO
\end{bNiceMatrix}.
\end{equation}
The scattering matrix $\mS^{\G_{12}}$ of the supersystem is constructed from the scattering matrices of the independent subsystems as

\begin{equation}
\setlength\arraycolsep{2pt}
\mS^{\G_{12}} = \mathrm{blockdiag}\left\{\mS^{\G_1}, \mS^{\G_2}\right\} =
\begin{bNiceMatrix}[first-row, first-col]
            & \iN                   & \iC_1 & \iC_2            \\
\iN         & \Block{2-2}{\mS^{\G_1}} &       & \mO              \\
\iC_1       &                       &       & \mO              \\
\iC_2       & \mO                   & \mO   & \mS^{\G_2}
\end{bNiceMatrix}.
\end{equation}

Therefore, by analogy with Eq.~(\ref{eq:matrix_loading}), the connection between $\G_{12}$ and $\Delta_{\G_1}^{\G_2}$ yields 
\begin{equation}
    \tilde{\mS}^{\G_{12}} = \mS^{\G_{12}}_{\iN\iN} + \mS^{\G_{12}}_{\iN\iC}
    \left(\left(\mS^{\Delta_{\G_1}^{\G_2}}\right)^{-1} - \mS^{\G_{12}}_{\iC\iC}\right)^{-1}
    \mS^{\G_{12}}_{\iC\iN},
    \label{eq5}
\end{equation}
where all the connected ports of the supersystem $\mathrm{G}_{12}$ are collected in $\iC=\iC_1 \cup \iC_2$.
By remarking that $\mS^{\Delta_{\G_1}^{\G_2}}$ is unitary and symmetric, and therefore equal to its inverse, we have
\begin{equation}
\setlength\arraycolsep{2pt}
\tilde{\mS}^{\G_{12}}
= \mS^{\G_1}_{\iN\iN}
+ \begin{bmatrix} \mS^{\G_{12}}_{\iN\iC_1} & \mO \end{bmatrix}
\begin{bmatrix} -\mS^{\G_1}_{\iC_1\iC_1} & \mI \\
                \mI & -\mS^{\G_2} \end{bmatrix}^{-1}
\begin{bmatrix} \mS^{\G_1}_{\iC_1\iN} \\ \mO \end{bmatrix}.
\label{eq:innerouter-b}
\end{equation}
The proof that $\tilde{\mS}^{\G_{12}} = \mS^{\G}$ is then obtained by simplifying this expression using the specific blockwise matrix inversion provided in Appendix~\ref{AppendixD}.

\begin{rem}
    Although Eq.\,(\ref{eq:innerouter-b}) ultimately yields the same result as Eq.\,(\ref{eq:innerouter-a}), Eq.\,(\ref{eq:innerouter-b}) comes with a computational drawback compared to Eq.\,(\ref{eq:innerouter-a}). Indeed, the inclusion of $\delta$-connections increases the number of connections mathematically included in the inversion problem. This is reflected in the fact that the size of the matrix to be inverted is doubled in Eq.\,(\ref{eq:innerouter-b}) compared to Eq.\,(\ref{eq:innerouter-a}). Nonetheless, it sets the stage for the derivation of a generic method applicable to arbitrary complex connections.
\label{rem:delta_connection_cost}
\end{rem}

\subsection{Generic supersystem connection}
\label{sec:generic_supersystem_connection}

The interpretation of Eq.\,(\ref{eq5}) allows us to write the generic form of the scattering matrix $\tilde{\mS}^{\super}$ of an internally connected supersystem as
\begin{equation}
    \tilde{\mS}^{\super} = \mS^{\super}_{\iN\iN} + \mS^{\super}_{\iN\iC} \left(\left(\mS^{\super}_\con\right)^{-1} - \mS^{\super}_{\iC\iC}\right)^{-1} \mS^{\super}_{\iC\iN},
\label{eq:generic}
\end{equation}
where $\mS^{\super}$ is the scattering matrix of the supersystem and $\mS^{\super}_\con$ is the scattering matrix of the system of connections made between the supersystem's ports $\iC$, the remaining free ports of the supersystem being $\iN$.

\begin{rem}
    It is possible to use any arbitrary connection scheme, provided the scattering matrix $\mS^{\super}_\con$ of the connections is obtainable. For lossless connections, the scattering matrix of the system of connections is unitary; $\left({\mS^{\super}_\con}\right)^\dagger \mS^{\super}_\con = \mathbf{I}$; thus, the inverse of $\mS^{\super}_\con$ is simply equal to its transpose conjugate: $\left({\mS^{\super}_\con}\right)^{-1} = \left({\mS^{\super}_\con}\right)^\dagger$. A special case thereof is a connection system only involving $\delta$-connections such that $\mathbf{S}_\mathrm{con}$ is a symmetric permutation matrix and hence equal to its inverse. 
\label{rem:unitarity}
\end{rem}
\begin{rem}
    Formulas derived from Eq.\,(\ref{eq:generic}) throughout this paper include a term of the form $\left( \left({\mS^{\super}_\con}\right)^{-1} - \mS^{\super}_{\iC\iC} \right)^{-1}$ involving two matrix inversions. In important special cases described in Remark~\ref{rem:unitarity}, the evaluation of $\left({\mS^{\super}_\con}\right)^{-1}$ is not computationally costly. Otherwise, it may be computationally advantageous to rewrite the term $\left( \left({\mS^{\super}_\con}\right)^{-1} - \mS^{\super}_{\iC\iC} \right)^{-1}$ as ${\mS^{\super}_\con}\left(\mI-\mS^{\super}_{\iC\iC}{\mS^{\super}_\con}\right)^{-1}$ or $\left(\mI-{\mS^{\super}_\con}\mS^{\super}_{\iC\iC}\right)^{-1}{\mS^{\super}_\con}$. Thereby, only one matrix inversion is required, at the expense of two additional matrix products.
\label{rem:inversion}
\end{rem}
\begin{rem}
     It is usually computationally more efficient to obtain the product $\mathbf{A}^{-1}\mathbf{B}$ by solving $\mathbf{A}\mathbf{X}=\mathbf{B}$ for $\mathbf{X}$ instead of performing the inversion and the matrix product sequentially. The speedup increases with the sparsity of $\mathbf{A}$. This can be used alongside Remark \ref{rem:inversion} and to partially compensate the additional computational cost involved by the generic connection, if the connection matrix is sparse (as it is the case for $\delta$-connections).
 \label{rem:solving}
\end{rem}

\subsection{Redheffer star product for connection schemes leaving free ports in both connected systems}
\label{sec:redheffer}

\begin{figure}[h]
    \centering
    \includegraphics[width=0.9\columnwidth]{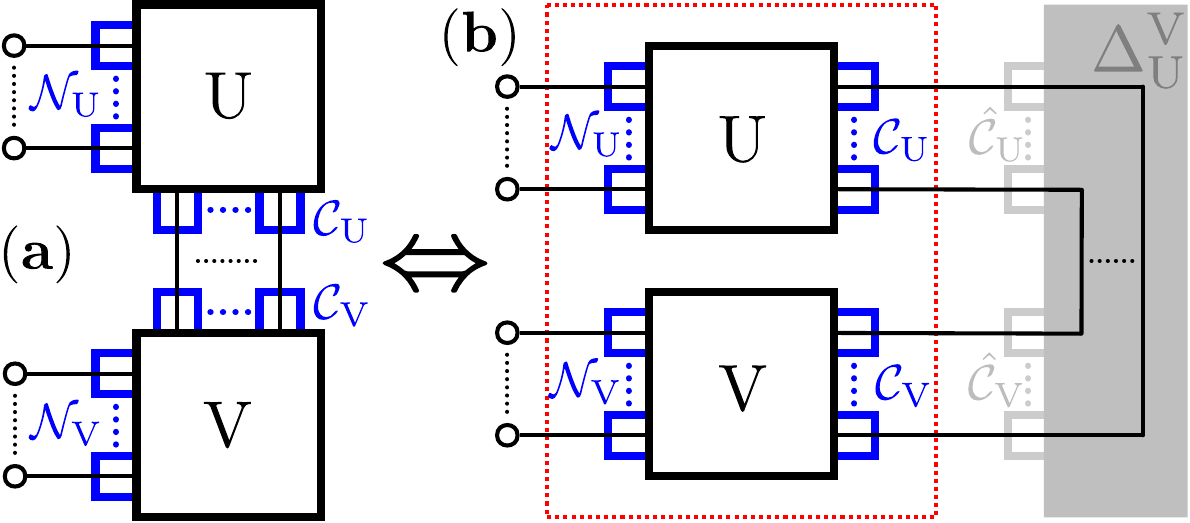}
    \caption{Cascade of two systems U and V with open ports after the connection. Blue rectangles represent sets of multiple ports in compacted depiction. (a) Conventional representation. (b) Representation to solve the problem using the cascade loading result from Fig.~\ref{fig:innnerouterequivalency} by identifying a supersystem (red-dotted frame) and connection system $\Delta_\U^\V$ composed of a set of $\delta$-connections (grey overlay).}
    \label{fig:genericcascade}
\end{figure}

We now extend the problem studied in Fig.\,\ref{fig:innnerouterequivalency} to the connection of two systems which both have free ports after being connected to each other~\cite{yang_new_2018}. As depicted in Fig.\,\ref{fig:genericcascade}(a), we refer to the systems as $\U$ and $\V$, to the sets of indices of their respective connected ports as $\iC_\U$ and $\iC_\V$ (satisfying $n(\iC_\U)=n(\iC_\V)$), and to the sets of indices of their respective free ports as $\iN_\U$ and $\iN_\V$.

To tackle this problem with the generic supersystem approach based on cascade loading from the previous subsection, we define a supersystem comprising $\U$ and $\V$ as well as a connection system $\Delta_\U^\V$, as depicted in Fig.\,\ref{fig:genericcascade}(b).
The scattering matrix of the supersystem is 
\begin{equation}
\mS^{\U\V} = \mathrm{blockdiag}\left\{\mS^\U,\mS^\V\right\} =
\begin{bNiceMatrix}[first-row, first-col]
        & \iP_\U & \iP_\V \\
\iP_\U  & \mS^\U & \mO    \\
\iP_\V  & \mO    & \mS^\V 
\end{bNiceMatrix},
\end{equation}
where 
\begin{equation}
    \iP_\U=\iN_\U\cup\iC_\U, \quad \iP_\V=\iN_\V\cup\iC_\V,
\end{equation}
and the scattering matrix of the connection system is
\begin{equation}
\mS_\con^{\U\V} = \mS^{\Delta_\U^\V} = 
\begin{bNiceMatrix}[first-row, first-col]
        & \hat{\mathcal{C}}_\U & \hat{\mathcal{C}}_\V \\
\hat{\mathcal{C}}_\U  & \mO    & \mI    \\
\hat{\mathcal{C}}_\V  & \mI    & \mO
\end{bNiceMatrix}.
\label{eq:single_cascade_def}
\end{equation}

We now re-partition the supersystem matrix by collecting free ports in $\iN=\iN_\U\cup\iN_\V$ and $\delta$-connected ports in $\iC=\iC_\U\cup\iC_\V$, yielding
\begin{equation}
\setlength\arraycolsep{0pt}
\begin{array}{cc}
\mS^{\U\V}_{\iN\iN} = \begin{bmatrix} \mS^\U_{\iN_\U\iN_\U} & \mO \\
                                        \mO & \mS^\V_{\iN_\V\iN_\V} \end{bmatrix},
& \ \mS^{\U\V}_{\iN\iC} = \begin{bmatrix} \mS^\U_{\iN_\U\iC_\U} & \mO \\
                                        \mO & \mS^\V_{\iN_\V\iC_\V} \end{bmatrix}, \\
\mS^{\U\V}_{\iC\iN} = \begin{bmatrix} \mS^\U_{\iC_\U\iN_\U} & \mO \\
                                        \mO & \mS^\V_{\iC_\V\iN_\V} \end{bmatrix},
& \mS^{\U\V}_{\iC\iC} = \begin{bmatrix} \mS^\U_{\iC_\U\iC_\U} & \mO \\
                                        \mO & \mS^\V_{\iC_\V\iC_\V} \end{bmatrix}.
\end{array}
\label{eq:partitioning_UV}
\end{equation}

Applying Eq.\,(\ref{eq:generic}) to determine the scattering matrix $\tilde{\mS}^\mathrm{UV}$ of the connected supersystem yields
\begin{equation}
\setlength\arraycolsep{0pt}
\begin{split}
& \tilde{\mS}^{\U\V} =
\begin{bmatrix} \mS^\U_{\iN_\U\iN_\U} & \mO                   \\
                \mO                   & \mS^\V_{\iN_\V\iN_\V} \end{bmatrix} + \\ & \!  
\begin{bmatrix} \mS^\U_{\iN_\U\iC_\U} & \mO                   \\
                \mO                   & \mS^\V_{\iN_\V\iC_\V} \end{bmatrix} \!
\begin{bmatrix} -\mS^\U_{\iC_\U\iC_\U} & \mI                   \\
                \mI                    & -\mS^\V_{\iC_\V\iC_\V} \end{bmatrix}^{-1} \!
\begin{bmatrix} \mS^\U_{\iC_\U\iN_\U} & \mO \\
                \mO                   & \mS^\V_{\iC_\V\iN_\V} \end{bmatrix}.
\end{split}
\label{eq:single_cascade_exec}
\end{equation}
The full reduction of Eq.\,(\ref{eq:single_cascade_exec}) (using the specific blockwise matrix inversion provided in Appendix~\ref{AppendixD}) yields the following expressions for the four blocks of $\tilde{\mS}^\mathrm{UV}$:
\begin{equation}
\begin{split}
& \tilde{\mS}^{\U\V}_{\iN_\U\iN_\U} = \mS^\U_{\iN_\U\iN_\U} - \mS^\U_{\iN_\U\iC_\U} \mS^\V_{\iC_\V\iC_\V} \mathbf{X}^{\U\V} \mS^\U_{\iC_\U\iN_\U},\\
& \tilde{\mS}^{\U\V}_{\iN_\U\iN_\V} = -\mS^\U_{\iN_\U\iC_\U} \mathbf{X}^{\V\U} \mS^\V_{\iC_\V\iN_\V},\\
& \tilde{\mS}^{\U\V}_{\iN_\V\iN_\U} = -\mS^\V_{\iN_\V\iC_\V} \mathbf{X}^{\U\V} \mS^\U_{\iC_\U\iN_\U},\\
& \tilde{\mS}^{\U\V}_{\iN_\V\iN_\V} = \mS^\V_{\iN_\V\iN_\V} - \mS^\V_{\iN_\V\iC_\V} \mS^\U_{\iC_\U\iC_\U} \mathbf{X}^{\V\U} \mS^\V_{\iC_\V\iN_\V},
\end{split}
\label{eq:redheffer}
\end{equation}
where 
\begin{equation}
\begin{split}
 & \mathbf{X}^{\U\V} = \left( \mS^\U_{\iC_\U\iC_\U} \mS^\V_{\iC_\V\iC_\V} - \mI \right)^{-1}, \\
& \, \mathbf{X}^{\V\U} = \left( \mS^\V_{\iC_\V\iC_\V} \mS^\U_{\iC_\U\iC_\U} - \mI \right)^{-1}.
\end{split}
\label{eq:redheffer_aux}
\end{equation}
The result from Eq.\,(\ref{eq:redheffer}) is widely used in the literature~\cite{simpson_generalized_1981, chu_generalized_1986, overfelt1989alternate, narayan_novel_2011, rumpf_improved_2011, yang_new_2018, zaky_fully_2021}) and known as the ``Redheffer star product'' \cite{redheffer_inequalities_1959}, which provides a compact notation to summarize Eqs.\,(\ref{eq:redheffer},\ref{eq:redheffer_aux}). Explicitly addressing the connection between ports $\iC_\U$ and $\iC_\V$, we write
\begin{equation}
    \tilde{\mS}^{\U\V} = \mS^\U \underset{\iC_\U,\iC_\V}{\star} \mS^\V.
\label{eq:redheffer_star}
\end{equation}

Of course, the equivalency of the connection of $\U$ and $\V$ with the connection of ${\U\V}$ and $\Delta_\U^\V$ can now also be stated in terms of Redheffer star products:
\begin{equation}
    \mS^\U \underset{\iC_\U,\iC_\V}{\star} \mS^\V \; = \; \mS^{\U\V} \underset{\iC_\U\cup\iC_\V,\hat{\mathcal{C}}_\U\cup\hat{\mathcal{C}}_\V}{\star} \mS^{\Delta_\U^\V}.
\end{equation}

\begin{rem}
We have derived the Redheffer star product starting from the cascade loading result in this section. However, one can also derive the cascade loading result as a special case of the Redheffer star product when $\iN_\V=\varnothing$. In this case, the loading formula requires us to invert a matrix of size $n\left(\iC_\U\right) \times n\left(\iC_\U\right)$, while the Redheffer star product requires us to invert a matrix of size $\left(n\left(\iC_\U\right)+n\left(\iC_\V\right)\right)\times\left(n\left(\iC_\U\right)+n\left(\iC_\V\right)\right)$, which is actually decomposed into two inversions of matrices of size $n\left(\iC_\U\right)\times n\left(\iC_\U\right)$ (note that $n\left(\iC_\U\right) =n\left(\iC_\V\right)$): the computation of $\XUV$ and $\XVU$. This reflects the physical fact that the interactions have to be resolved as viewed from both sides of the connection. This also implies that if the systems $\U$ and $\V$ are both reciprocal, at least as viewed from their connected ports (i.e., $\mS_{\iC_\U\iC_\U}^{\U} = \left(\mS_{\iC_\U\iC_\U}^\U\right)^T$ and $\mS_{\iC_\V\iC_\V}^\V = \left(\mS_{\iC_\V\iC_\V}^\V\right)^T$), then the interactions from both sides of the connection are similar (i.e., $\mathbf{X}^{\V\U}=\left(\mathbf{X}^{\U\V}\right)^T$) and only one matrix inversion is required.
\end{rem}

\section{Global method for \\arbitrarily complex connections}
\label{sec:global}

In this section, we illustrate a global (i.e., non-iterative and closed-form) method to evaluate the scattering matrix resulting from arbitrarily complex connections between multi-port systems, without any limitations on the types of connections.  
The mathematical structure of this global method is clear and interpretable. These benefits initially come at the expense of a larger algebraic problem size, and thus an increased computational cost. However, leveraging the inner-outer equivalency can subsequently limit the algebraic problem size. This method, denoted \emph{reducibility}, is detailed in Sec.~\ref{sec:reducibility}. The key advantage of the global method leveraged in our present paper is the \emph{updatability} of previously solved related versions of the problem, as detailed in Sec.~\ref{sec:updatability}. A final advantage is the possibility to perform a global (non-iterative) \emph{recovery of power waves traveling through the connected ports}, as detailed in Sec.~\ref{sec:generalizedconnect_potentials}. Reducibility, updatability, and power wave recovery techniques can be implemented in combination.

\subsection{Principle of global method}

The key steps of the global method rely on the generic connection method described in Secs. \ref{sec:connections_as_scattering} and \ref{sec:generic_supersystem_connection}  and are summarized as follows:
\begin{enumerate}[label=(\roman*)]
    \item All involved constituent scattering systems are gathered into a supersystem.
    \item The sets of indices of the free ports and connected ports of the involved scattering matrices are identified.
    \item A connection system comprising the inner $\delta$-connections of the supersystem is defined.
    \item The scattering matrix of the connected supersystem is obtained by applying Eq.\,(\ref{eq:generic}).
\end{enumerate}

\subsection{Illustration with meta-network as prototypical generalized connection scheme}
\label{sec:meta-network_example}

\begin{figure}[h]
    \centering
    \includegraphics[width=\columnwidth]{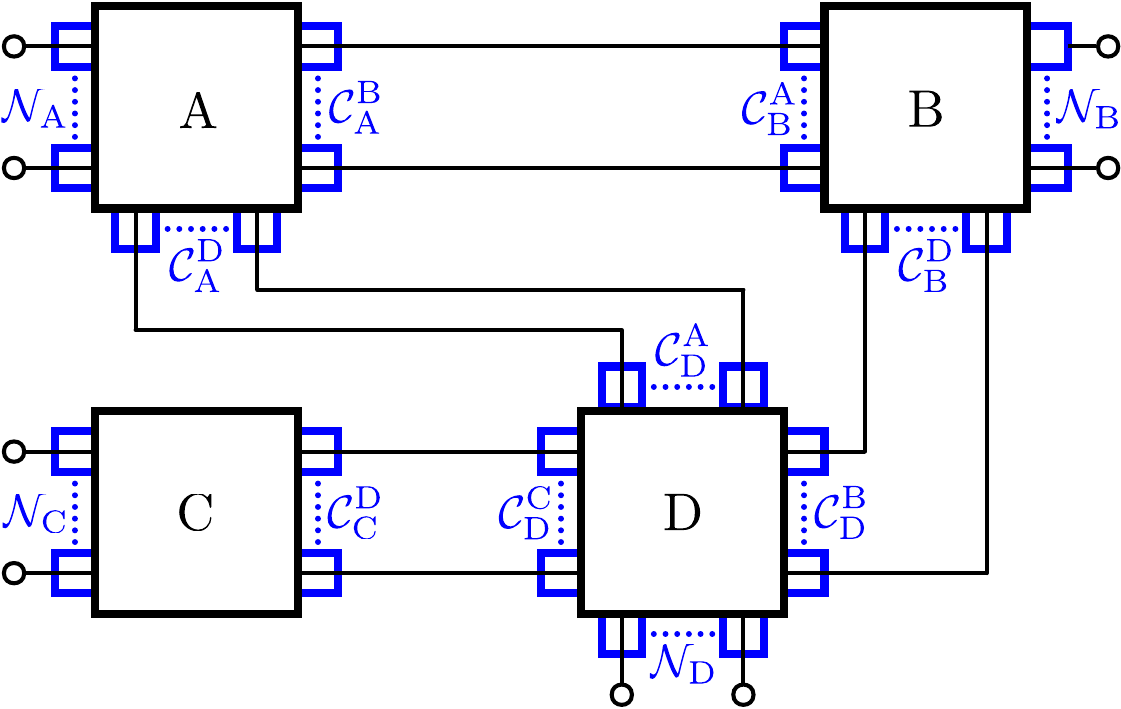}
    \caption{Meta-network example comprising serial, parallel and cyclic connections between the four scattering systems A, B, C and D.}
    \label{fig:meta-network}
\end{figure}

The topology of connection schemes for scattering systems can exhibit diverse characteristics. The connection patterns may involve parallel junctions, serial chains, or cycles. To illustrate our approach with a prototypical connection scheme incorporating all of these connection patterns, we focus on the one illustrated in Fig.\,\ref{fig:meta-network}. Because the systems A, B, C and D are themselves networks, we refer to this network of networks as ``meta-network''.

\begin{figure}[h]
    \centering
    \includegraphics[width=0.65\columnwidth]{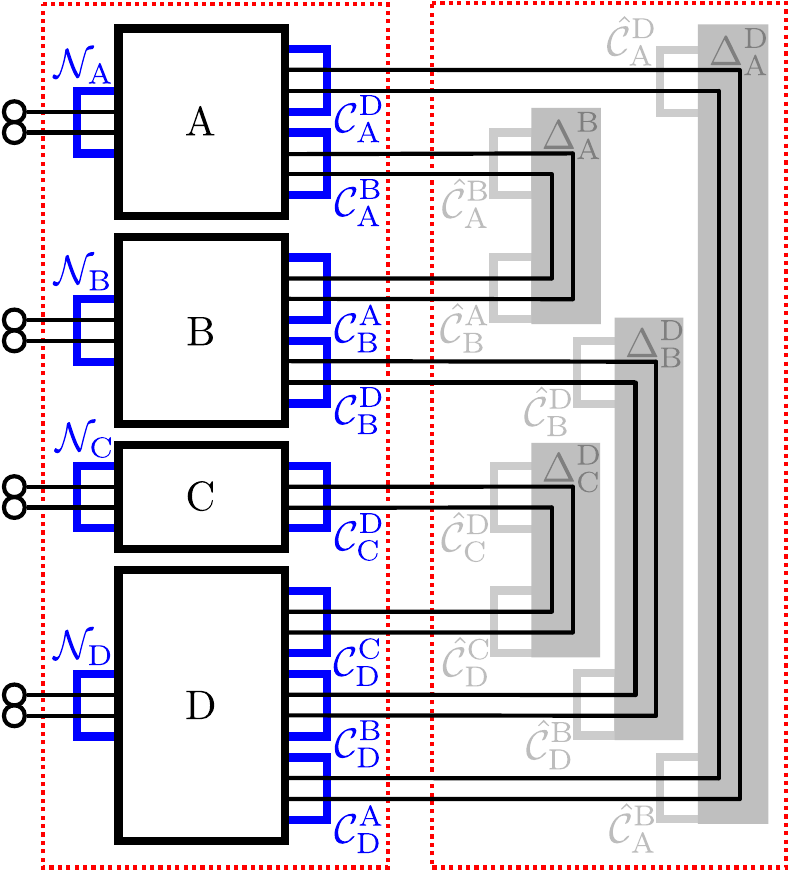}
    \caption{Schematic of the application of the global method to the meta-network from Fig.\,\ref{fig:meta-network}. The ports are depicted in a more compact manner than in Fig.\,\ref{fig:meta-network}. Red-dotted frames identify the supersystem and the connection system. Grey overlays identify sets of $\delta$-connections.}
\label{fig:meta-network_global_connection}
\end{figure}

To start, in step (i), we define the scattering matrix $\mS^\mathrm{ABCD}$ of the supersystem gathering the scattering matrices of the subsystems $\A$, $\B$, $\C$, $\D$, as illustrated by the left red-dotted frame in Fig.\,\ref{fig:meta-network_global_connection}:
\begin{equation}
\setlength\arraycolsep{4pt}
\begin{split}
\mS^\mathrm{ABCD} & = \mathrm{blockdiag}\left\{\mS^\A,\mS^\B,\mS^\C,\mS^\D\right\}, \\
            & = \begin{bNiceMatrix}[first-col, first-row]
            & \iP_\A & \iP_\B & \iP_\C & \iP_\D  \\
\iP_\A      & \mS^\A  & \mO    & \mO    & \mO    \\
\iP_\B      & \mO     & \mS^\B & \mO    & \mO    \\
\iP_\C      & \mO     & \mO    & \mS^\C & \mO    \\
\iP_\D      & \mO     & \mO    & \mO    & \mS^\D
\end{bNiceMatrix}, 
\end{split}
\label{eq:meta-network_S_super}
\end{equation}
where 
\begin{equation}
\begin{array}{ll}
\iP_\A = \iN_\A \cup \iC_\A^\B \cup \iC_\A^\D,                &
\iP_\B = \iN_\B \cup \iC_\B^\A \cup \iC_\B^\D,                \\
\iP_\C = \iN_\C \cup \iC_\C^\D,                               &
\iP_\D = \iN_\D \cup \iC_\D^\A \cup \iC_\D^\B \cup \iC_\D^\C.
\end{array}
\end{equation}

Then, in step (ii), we partition the supersystem scattering matrix $\mS^\mathrm{ABCD}$ by collecting the free ports in $\iN$ and the connected ports in $\iC$, where
\begin{equation}
\begin{array}{l}
\iN = \iN_\A \cup \iN_\B \cup \iN_\C \cup \iN_\D, \\
\iC = \iC_\A^\B \cup \iC_\A^\D \cup \iC_\B^\A \cup \iC_\B^\D \cup \iC_\C^\D \cup \iC_\D^\A \cup \iC_\D^\B \cup \iC_\D^\C.
\end{array}
\end{equation}
The matrices $\mS^\mathrm{ABCD}_{\iN\iN}$, $\mS^\mathrm{ABCD}_{\iN\iC}$, $\mS^\mathrm{ABCD}_{\iC\iN}$ and $\mS^\mathrm{ABCD}_{\iC\iC}$ are block-diagonal matrices that can all be expressed as
\begin{equation}
    \mS^\mathrm{ABCD}_{\iI\iJ} = \mathrm{blockdiag}\left\{
    \mS^\A_{\{\iI\iJ\}},\mS^\B_{\{\iI\iJ\}},\mS^\C_{\{\iI\iJ\}},\mS^\D_{\{\iI\iJ\}}
    \right\},
\label{eq:submatrix_selection}
\end{equation}
where we have introduced a specific notation for indexing the scattering matrix of any subsystem $\X \in \{\mathrm{A,B,C,D}\}$ over any couple of port index sets $(\iI, \iJ)\in\{\iN,\iC\}^2$ to declutter the notation of the partitioning involved in Eq.\,(\ref{eq:partitioning_UV}):
\begin{equation}
\mS^\X_{\{\iI\iJ\}} = \mS^\X_{\iI\cap\iP_\X\,\iJ\cap\iP_\X}.
\end{equation}

Next, in step (iii), we define the scattering matrix of the connection system comprising all the $\delta$-connections, illustrated by the right red-dotted frame in Fig.\,\ref{fig:meta-network_global_connection}. The connection system's scattering matrix is constructed by, first, gathering the scattering matrices of the sets of $\delta$-connections between pairs of systems,
\begin{equation}
    \widehat{\mS^\mathrm{ABCD}_\con} = \mathrm{blockdiag}\left\{\mS^{\Delta_\A^\B},\mS^{\Delta_\A^\D},\mS^{\Delta_\B^\D},\mS^{\Delta_\C^\D}\right\},
\end{equation}
and, second, re-ordering $\widehat{\mS^\mathrm{ABCD}_\con}$ to match the order of $\iC$:
\begin{equation}
\setlength\arraycolsep{4pt}
\mS_\con^\mathrm{ABCD} = 
\begin{bNiceMatrix}[first-col, first-row]
            & \hat{\mathcal{C}}_\A^\B & \hat{\mathcal{C}}_\A^\D & \hat{\mathcal{C}}_\B^\A & \hat{\mathcal{C}}_\B^\D & \hat{\mathcal{C}}_\C^\D & \hat{\mathcal{C}}_\D^\A & \hat{\mathcal{C}}_\D^\B & \hat{\mathcal{C}}_\D^\C  \\
\hat{\mathcal{C}}_\A^\B   & \mO       & \mO       & \mI       & \mO       & \mO       & \mO       & \mO        & \mO       \\
\hat{\mathcal{C}}_\A^\D   & \mO       & \mO       & \mO       & \mO       & \mO       & \mI       & \mO        & \mO       \\
\hat{\mathcal{C}}_\B^\A   & \mI       & \mO       & \mO       & \mO       & \mO       & \mO       & \mO        & \mO       \\
\hat{\mathcal{C}}_\B^\D   & \mO       & \mO       & \mO       & \mO       & \mO       & \mO       & \mI        & \mO       \\
\hat{\mathcal{C}}_\C^\D   & \mO       & \mO       & \mO       & \mO       & \mO       & \mO       & \mO        & \mI       \\
\hat{\mathcal{C}}_\D^\A   & \mO       & \mI       & \mO       & \mO       & \mO       & \mO       & \mO        & \mO       \\
\hat{\mathcal{C}}_\D^\B   & \mO       & \mO       & \mO       & \mI       & \mO       & \mO       & \mO        & \mO       \\
\hat{\mathcal{C}}_\D^\C   & \mO       & \mO       & \mO       & \mO       & \mI       & \mO       & \mO        & \mO
\end{bNiceMatrix}.
\label{eq:single_cascade_def}
\end{equation}

Finally, in step (iv), we obtain the scattering matrix of the connected supersystem by applying Eq.\,(\ref{eq:generic}):
\begin{multline}
    \tilde{\mS}^\mathrm{ABCD} = \\ \mS^\mathrm{ABCD}_{\iN\iN} + \mS^\mathrm{ABCD}_{\iN\iC}
    \left( \left(\mS_\con^\mathrm{ABCD}\right)^{-1} - \mS^\mathrm{ABCD}_{\iC\iC} \right)^{-1}
    \mS^\mathrm{ABCD}_{\iC\iN}.
\label{eq:meta-network-cascade-global}
\end{multline}
Since $\mS_\con^\mathrm{ABCD}$ is a symmetric permutation matrix, it equals its inverse, as pointed out in Remark~\ref{rem:unitarity}. Thus, the matrix ${\left(\mS_\con^\mathrm{ABCD}\right)^{-1} - \mS^\meta_{\iC\iC}}$ which has to be inverted in Eq.\,(\ref{eq:meta-network-cascade-global}) takes the following form
\newcommand{\SaCC}{\ensuremath{\mS^\A_{\{\iC\iC\}}}}
\newcommand{\SbCC}{\ensuremath{\mS^\B_{\{\iC\iC\}}}}
\newcommand{\ScCC}{\ensuremath{\mS^\C_{\{\iC\iC\}}}}
\newcommand{\SdCC}{\ensuremath{\mS^\D_{\{\iC\iC\}}}}
\begin{equation}
\setlength\arraycolsep{4pt}
\begin{split}
& {\left(\mS_\con^\mathrm{ABCD}\right)^{-1}} - \mS^\mathrm{ABCD}_{\iC\iC}  = \\
& \begin{bNiceMatrix}[first-col, first-row]
            & \iC_\A^\B & \iC_\A^\D & \iC_\B^\A & \iC_\B^\D & \iC_\C^\D & \iC_\D^\A & \iC_\D^\B & \iC_\D^\C  \\
\iC_\A^\B   & \Block{2-2}{-\SaCC}&  & \mI       & \mO       & \mO       & \mO       & \mO        & \mO       \\
\iC_\A^\D   &           &           & \mO       & \mO       & \mO       & \mI       & \mO        & \mO       \\
\iC_\B^\A   & \mI       & \mO       & \Block{2-2}{-\SbCC}&  & \mO       & \mO       & \mO        & \mO       \\
\iC_\B^\D   & \mO       & \mO       &           &           & \mO       & \mO       & \mI        & \mO       \\
\iC_\C^\D   & \mO       & \mO       & \mO       & \mO       & -\ScCC    & \mO       & \mO        & \mI       \\
\iC_\D^\A   & \mO       & \mI       & \mO       & \mO       & \mO       & \Block{3-3}{-\SdCC}&   &           \\
\iC_\D^\B   & \mO       & \mO       & \mO       & \mI       & \mO       &           &            &           \\
\iC_\D^\C   & \mO       & \mO       & \mO       & \mO       & \mI       &           &            & 
\end{bNiceMatrix}.
\end{split}
\label{eq:meta-network-global-inverse}
\end{equation}

\section{Reducibility}
\label{sec:reducibility}

The global method presented in Sec.~\ref{sec:global}  constitutes a closed-form, transparent approach for evaluating arbitrarily complex connections of a multitude of scattering systems. However, it is usually not the computationally most efficient approach, as already seen with the simple examples in Sec.~\ref{sec:connections_as_scattering} and Sec.~\ref{sec:redheffer}. 
Indeed, the global method requires by construction the inversion of the $n(\iC)\times n(\iC)$ matrix $\left(\left(\mS^\mathrm{ABCD}_\con\right)^{-1} -\mS^\mathrm{ABCD}_{\iC\iC}\right)$, as seen in Eq.\,(\ref{eq:meta-network-cascade-global}), which typically dominates the computational cost.\footnote{Based on the number of arithmetic operations, we consider that the computational complexity for the inversion of an $n \times n$ matrix is $\mathcal{O}\left(n^3\right)$.~\cite{farebrother2018linear}} The matrix to be inverted typically contains many zero blocks, as seen in Eq.\,(\ref{eq:meta-network-global-inverse}).

In this section, we demonstrate a reducibility property of the global method based on the inner-outer equivalency which can be applied to significantly improve the computational efficiency while maintaining the advantages of a closed-form, transparent and non-iterative approach.
In terms of computational efficiency, exploiting the reducibility property provides a non-iterative middle ground between the non-iterative global method from Sec.~\ref{sec:global} and iterative methods that sequentially apply Eq.\,(\ref{eq:matrix_loading}) (cascade loading) or Eq.\,(\ref{eq:redheffer_star}) (Redheffer star product), adding on one subsystem after the other.

\subsection{Principle}

The global method from Sec.~\ref{sec:global} gathers all subsystems comprised in a supersystem which then features only inner $\delta$-connections. Given the inner-outer equivalency, it is alternatively possible to treat some of the subsystems as connections rather than as part of the supersystem. 
This will reduce the number of $\delta$-connections and, consequently, the size of the matrix to be inverted, as pointed out in Remark \ref{rem:delta_connection_cost}.

There are usually multiple ways to perform the reduction. If multiple systems are interpreted as constituting connections rather than being included in the supersystem, then the systems treated as connections must not be directly connected to each other.\footnote{In principle, it is also possible to treat two directly connected systems as connections by inserting an auxiliary system composed of $\delta$-connections into the supersystem, but such an approach does not reduce the size of the supersystem and is hence unnecessarily complicated.} For concreteness, consider the example of our meta-network from Fig.~\ref{fig:meta-network}: A, B or D could be treated as connections; however, only one of them can be treated as part of the connection system in any given reduction because all three are directly connected to each other. We illustrate the reduction of the meta-network interpreting D as part of the connection system in Sec.~\ref{sec:reduction_meta-network}. 

A ``full'' reduction would decrease the dimensions of the supersystem by halving its number of connected ports, because for each pair of connected ports, only one would be part of the supersystem after reduction. 
After a full reduction, no $\delta$-connections remain. 
However, not all connection schemes can be fully reduced. Indeed, connection schemes involving connection cycles of an odd number of scattering systems, such as the cycle involving A, B and D in our meta-network, cannot be fully reduced.

Irrespective of whether the reduction is full or not, it results in considering an outer connection between two groups of systems (the connection system and the supersystem), each potentially involving free ports, such that the problem can be expressed in terms of the Redheffer star product. (This is in contrast to the global method from Sec.~\ref{sec:global} in which only the supersystem has free ports such that it constitutes an instance of cascade loading.) The reduction is hence still fully compatible with having a closed-form and non-iterative approach. In this regard, the present section can be viewed as casting related reduction concepts discussed in Ref.~\cite{yang_new_2018} as part of a recursive scheme into a non-iterative closed form.

\subsection{Reducibility of the meta-network}
\label{sec:reduction_meta-network}

\begin{figure}[h]
    \centering
    \includegraphics[width=0.75\columnwidth]{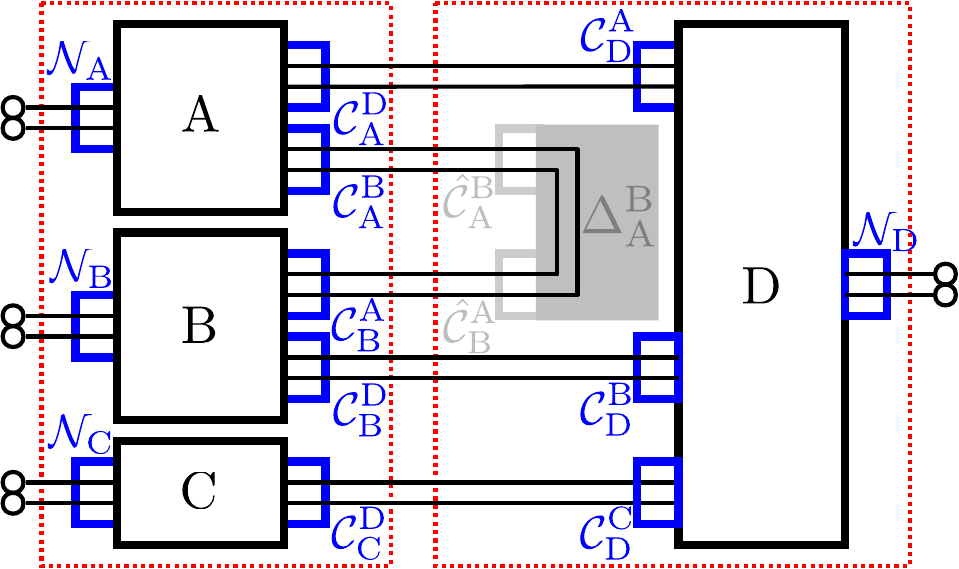}
    \caption{Schematic of applying the reducibility property of the global method to the meta-network from Fig.\,\ref{fig:meta-network}. The systems A, B and C are comprised in the supersystem (left red-dotted framing) whereas the system D is a part of the connection system (right red-dotted framing) along with the set of $\delta$-connections $\Delta_\A^\B$.}
    \label{fig:meta-network_reduced}
\end{figure}

As mentioned, the meta-network from Fig.\,\ref{fig:meta-network} is not fully reducible because it involves a cycle of an odd number of systems (A, B and D) but it is possible to treat the system D as part of the connection system rather than including it in the supersystem. This interpretation is illustrated in Fig.\,\ref{fig:meta-network_reduced} and results in the following scattering matrix for the supersystem:
\begin{equation}
\mS^\mathrm{ABC} = 
\begin{bNiceMatrix}[first-col, first-row]
            & \iP_\A & \iP_\B & \iP_\C   \\
\iP_\A      & \mS^\A  & \mO    & \mO        \\
\iP_\B      & \mO     & \mS^\B & \mO        \\
\iP_\C      & \mO     & \mO    & \mS^\C     \\
\end{bNiceMatrix}, 
\label{eq:meta-network_S_super_reduced}
\end{equation}
and the following scattering matrix for  the connection system:
\begin{equation}
\begin{split}
\mS_\con^\mathrm{ABC} & = \mathrm{blockdiag}\left\{\mS^{\Delta_\A^\B}, \mS^\D\right\} \\
                      & = \begin{bNiceMatrix}[first-col, first-row]
            & \hat{\mathcal{C}}_\A^\B & \hat{\mathcal{C}}_\B^\A  & \iC_\D^\A & \iC_\D^\B        & \iC_\D^\C & \iN_\D \\
\hat{\mathcal{C}}_\A^\B   & \mO       & \mI        & \mO       & \mO              & \mO       & \mO    \\
\hat{\mathcal{C}}_\B^\A   & \mI       & \mO        & \mO       & \mO              & \mO       & \mO    \\
\iC_\D^\A   & \mO       & \mO        & \Block{4-4}{\mS^\mathrm{D}}  &           &        \\
\iC_\D^\B   & \mO       & \mO        &           &                  &                    \\
\iC_\D^\C   & \mO       & \mO        &           &                  &                   \\
\iN_\D      & \mO       & \mO        &           &                  & 
\end{bNiceMatrix}.
\end{split}
\label{eq:single_cascade_def_reduced}
\end{equation}
As noted, the connection system now includes free ports (unlike in the global method from Sec.~\ref{sec:global}) such that obtaining the scattering matrix of the meta-network now requires us to apply the Redheffer star product from Eq.\,(\ref{eq:redheffer_star}):
\begin{equation}
    \tilde{\mS}^\mathrm{ABCD} = \mS^\mathrm{ABC} \underset{\iC^\mathrm{ABC}, \iC^\mathrm{ABC}_\con}{\star} \mS_\con^\mathrm{ABC},
\end{equation}
where $\iC^\mathrm{ABC}$ and $\iC^\mathrm{ABC}_\con$ collect in a matching order the ports of the supersystem and the connection system, respectively, that are connected:
\begin{equation}
\begin{array}{l}
    \iC^\mathrm{ABC} = \iC_\A^\B \cup \iC_\B^\A \cup \iC_\A^\D \cup \iC_\B^\D \cup \iC_\C^\D,\\
    \iC^\mathrm{ABC}_\con = \hat{\mathcal{C}}_\A^\B \cup \hat{\mathcal{C}}_\B^\A \cup \iC_\D^\A \cup \iC_\D^\B \cup \iC_\D^\C.\\
    \end{array}
\end{equation}
\begin{rem}
    Since the connection system's scattering matrix in the reduced case is block-diagonal, its inverse can be easily constructed based on the inverses of its blocks. Also, Remark \ref{rem:unitarity} can further alleviate the computational burden of its evaluation.
\end{rem}

If, however, D did not have any free ports, we could use the simpler cascade-loading Eq.\,(\ref{eq:generic}), akin to Sec.~\ref{sec:global}. To illustrate this, let us replace D by another system $\D^\prime$ without any free ports (i.e., $\iN_{\D^\prime}=\varnothing$) in the meta-network, yielding the ``modified meta-network''. The supersystem comprising A, B and C would remain unchanged. However, the connection system would take the following form
\begin{equation}
{\mS_\con^\mathrm{ABC}}^\prime = 
 \begin{bNiceMatrix}[first-col, first-row]
            & \hat{\mathcal{C}}_\A^\B & \hat{\mathcal{C}}_\B^\A  & \iC_\D^\A & \iC_\D^\B & \iC_\D^\C  \\
\hat{\mathcal{C}}_\A^\B   & \mO    & \mI          & \mO        & \mO      & \mO       \\
\hat{\mathcal{C}}_\B^\A   & \mI       & \mO           & \mO      & \mO        & \mO       \\
\iC_\D^\A   & \mO       & \mO             & \Block{3-3}{\mS^\mathrm{D^\prime}}&    &           \\
\iC_\D^\B   & \mO       & \mO            &           &            &           \\
\iC_\D^\C   & \mO       & \mO           &           &            & 
\end{bNiceMatrix}.
\label{eq:single_cascade_def_reduced_prime}
\end{equation}
The resulting scattering matrix of the modified meta-network would then be given by applying Eq.\,(\ref{eq:generic}):
\begin{equation}
    \tilde{\mS}^\mathrm{ABCD^\prime} = \mS^\mathrm{ABC}_{\iN\iN} + \mS^\mathrm{ABC}_{\iN\iC}
    \left( \left({\mS_\con^\mathrm{ABC}}^\prime \right)^{-1} - \mS^\mathrm{ABC}_{\iC\iC} \right)^{-1}
    \mS^\mathrm{ABC}_{\iC\iN}.
\label{eq:meta-network-reduced-connection}
\end{equation}

To systematically evaluate potential improvements in computational efficiency, we conducted an exhaustive study taking the meta-network from Fig.\,\ref{fig:meta-network} as example. Specifically, our study assumed that the scattering systems A, B, C and D (or $\D^\prime$) are graphs for which we can analytically evaluate the scattering matrix and the ground truth of the overall system (see Appendix~\ref{Appendix_graphs} for details). For simplicity, we used the same number $N_\mathrm{bus}$ of ports for each set of connections between two subsystems of the meta-network; the total number of ports was thus equal to $12N_\mathrm{bus}$. One node was assigned to each port at a random location within the domain $(x,y)\in\left[0,1\right]^2$ on a 2D plane and bonds were established by randomly selecting 50\% of all possible connections between nodes. The wavevector was set to $k=3+0.05\jmath$. Results were averaged on a number of graphs that was decreased as $N_\mathrm{bus}$ got larger, because of the self-averaging property as well as the higher computational cost for scenarios with higher values of $N_\mathrm{bus}$.

\begin{figure}[h]
    \centering
    \includegraphics[width=0.99\columnwidth]{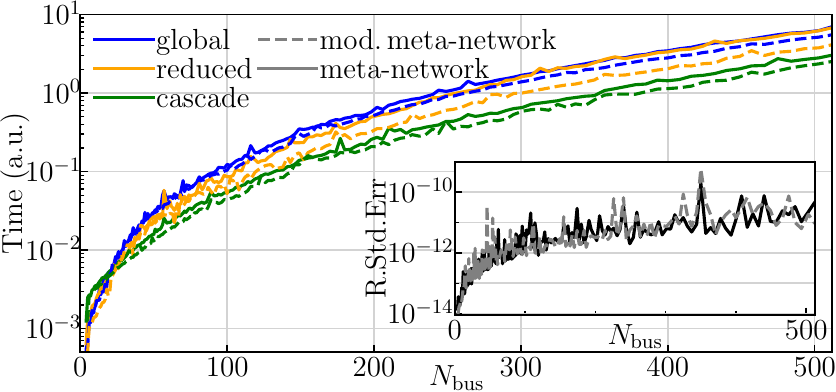}
    \caption{Computation time (main figure) and relative standard error (inset) for the evaluation of the scattering matrix for the meta-network from Fig.\,\ref{fig:meta-network} and the modified meta-network, using three distinct methods: global (Sec.~\ref{sec:global}), reduced (Sec.~\ref{sec:reducibility}), cascade (iterative application of the Redheffer star product, adding on one subsystem after the other). All evaluations are performed analytically by using graphs as scattering systems (see Appendix~\ref{Appendix_graphs} for details). $N_\mathrm{bus}$ is the number of ports for every set of connections between two subsystems. The relative standard error (R.Std.Err.) of the three methods are undistinguishable.}
    \label{fig:reducibility_benchmark}
\end{figure}

The computation time and relative standard error for a wide range of values of $N_\mathrm{bus}$ are summarized in Fig.\,\ref{fig:reducibility_benchmark} for the meta-network and the modified meta-network, for three methods of evaluating the resulting scattering matrix: global (Sec.~\ref{sec:global}), reduced (Sec.~\ref{sec:reducibility}) and cascade (iterative application of the Redheffer star product, adding on one subsystem after the other). All methods are seen to achieve comparable accuracy in the inset of Fig.\,\ref{fig:reducibility_benchmark}.
We observe that the iterative cascade method is roughly two times faster than the global method (because of the larger memory allocation delay of the global method) except for small values of $N_\mathrm{bus}$ (because the memory access latency piles up with each iteration). The computational improvement of the reduced method is noticeably more significant in the case of the modified meta-network because the computational cost of cascade loading [Eq.\,(\ref{eq:generic})] is smaller than the one of the Redheffer star product [Eq.~(\ref{eq:redheffer})].

\subsection{Reducibility of (de)multiplexing junctions and chain cascades}

For the sake of completeness, we end this section on reducibility by briefly discussing two simpler but practically important types of connections that are fully reducible.

\begin{figure}
    \centering
    \includegraphics[width=0.99\columnwidth]{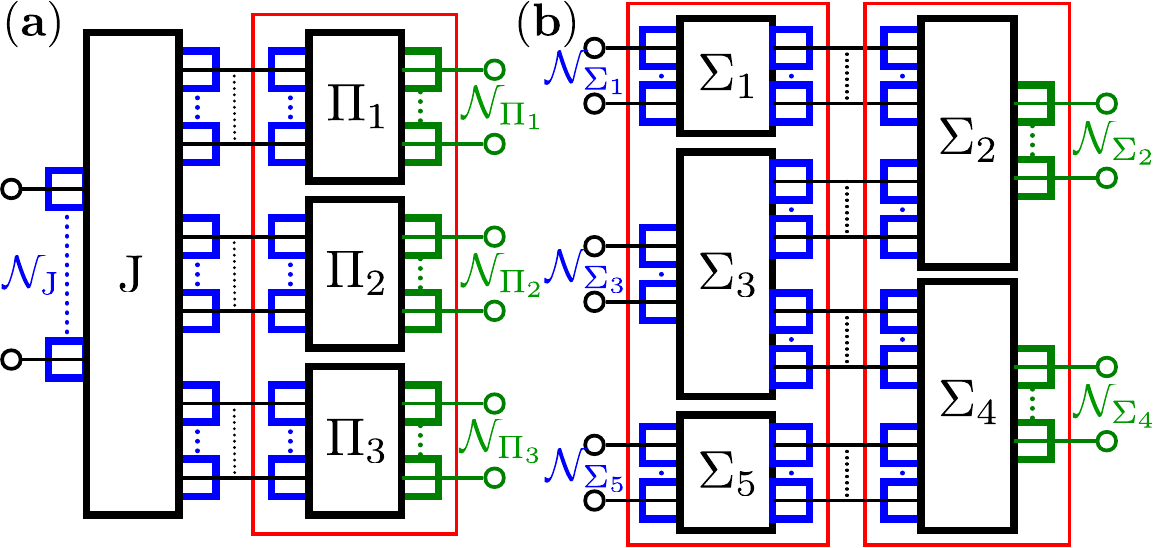}
    \caption{Reducibility of (a) a generic (de)multiplexing junction and (b) a generic chain cascade. The red boxes indicate convenient definitions of supersystems such that both problems can be treated with a single Redheffer star product. }
    \label{fig:reduction_junction_cascade}
\end{figure}

The first example is a generic (de)multiplexing junction, as illustrated in Fig.\,\ref{fig:reduction_junction_cascade}(a); special cases thereof have been considered in many works, including Refs.~\cite{anderson_cascade_1966, chu_generalized_1986, itoh_scattering_1995, shekel_junction_1974, csontos_scattering_matrix_2002,cao_applications_2015}. As highlighted by the red box in Fig.\,\ref{fig:reduction_junction_cascade}(a), the problem can easily be cast into the form of a Redheffer star product involving a joining system J and a supersystem comprising the incoming/outgoing parallel subsystems $\Pi_i$. 

This example also relates to RIS-parametrized radio environments. In the case of a conventional diagonal RIS, it specialized to a scenario in which the parallel subsystems $\Pi_i$ have no free ports and only one connected port each. In fact, this is then just a simple case of cascade loading. In the case of BD-RIS, as recently pointed out in Ref.~\cite{del2024physics}, the tunable load circuit terminating the auxiliary ports of the radio environment can be decomposed into a static circuit with auxiliary ports terminated by individual tunable loads. Hence, the connection between radio environment and the static part of the load circuit directly maps into the (de)multiplexing junction scenario where the number of junctions equals the number of groups of RIS elements in the BD-RIS (see Ref.~\cite{shen2021modeling} for details on the definition of group-connected RIS elements in BD-RIS). As highlighted in Ref.~\cite{del2024physics}, considering the resulting connected system and the termination of its auxiliary ports by individual loads allows one to directly apply algorithms developed for diagonal RIS to the realm of BD-RIS.

\begin{figure}[h]
    \centering
    \includegraphics[width=\columnwidth]{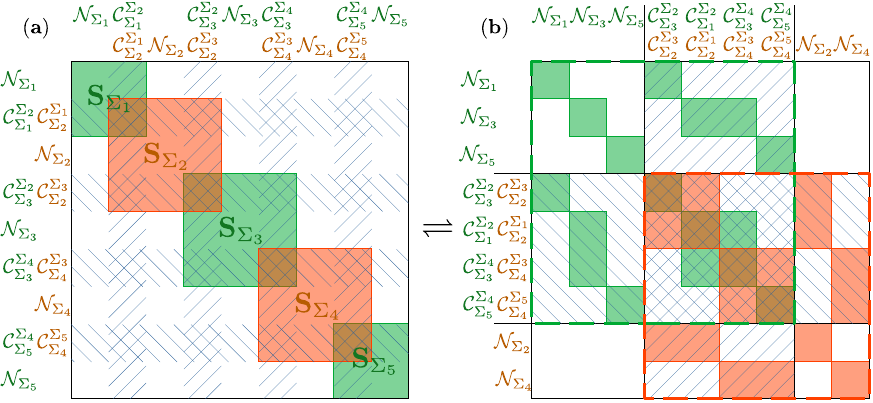}
    \caption{Illustration of the fully reduced evaluation of the generic chain cascade shown in  Fig.~\ref{fig:reduction_junction_cascade}(b) using the Redheffer star product: (a) canonical block-diagonal superimposed representation; (b) repartitioned superimposed representation analogous to Eq.~(\ref{eq:partitioning_UV}). White: null blocks. In (a), the green (resp. orange) blocks correspond to scattering matrices of subsystems with odd (resp. even) indices; the hatching identifies lines and columns associated with connected ports. In (b), each of these blocks is split into subblocks as part of the repartitioning analogous to Eq.~(\ref{eq:partitioning_UV}). In (b), the green-dashed (resp. red-dashed) frame identify the repartitioned scattering matrices of the supersystems comprising subsystems with odd (resp. even) indices. The black lines divide each repartitioned supersystem into the four blocks appearing in Eq.~(\ref{eq:partitioning_UV}). 
    The cross-hatching identifies the repartitioned supersystems' blocks which are involved in the two matrix inversions in Eq.~(\ref{eq:redheffer_aux}); these blocks are seen to be block-tridiagonal. The matrix inversions in Eq.~(\ref{eq:redheffer_aux}) are required for the evaluation of the Redheffer star product.
    }
    \label{fig:cascade_reordering}
\end{figure}

The second example is a generic chain cascade, again already the subject of several prior studies~\cite{yang_new_2018,de2022single}. We do \textit{not} assume that all connected systems have the same number of ports. As stated earlier, the chain cascade also appears in the physics-consistent description of stacked intelligent metasurfaces~\cite{SIM_clerckx}. We illustrate in Fig.\,\ref{fig:reduction_junction_cascade}(b) a chain of serially connected systems $\Sigma_i$. We also highlight in the figure that the problem can once again easily be cast into the form of a Redheffer star product. One simply indexes the subsystems of the chain cascade in order and interprets the chain cascade as connecting one supersystem $\left\{\Sigma_i \mid i \text{ is odd}\right\}$ comprising every subsystem with an odd index $i$ to another supersystem $\left\{\Sigma_i \mid i \text{ is even}\right\}$ comprising every remaining subsystem with an even index $i$. Incidentally, under the restriction that all systems have the same number of input and output ports, the chain cascade problem can also be tackled in the traditional way of conversions to ABCD or T matrices~\cite{speciale_projective_1981}, although it is already well documented that this traditional approach is computationally inefficient~\cite{de2022single}. The application of our fully reduced global method requires the gathering and block-partitioning of the involved scattering matrices as represented in Fig.\,\ref{fig:cascade_reordering}. It appears that the inversion problem translates into two  block-tridiagonal matrix inversions (unless the first or last system in the chain does not have any free ports in which case the Redheffer star product simplifies to cascade loading), as in Ref.~\cite{petersen_block_2008}, that can be recursively optimized using iterative blockwise inversion methods, as in Ref.~\cite{bachiller_efficient_2007}.

\section{Updatability}
\label{sec:updatability}

\newcommand{\Gl}{\ensuremath{\mathrm{\Gamma}}}
\newcommand{\E}{\ensuremath{\mathrm{E}}}
\newcommand{\mSi}{\mathbf{\overline{\mS}}}
\newcommand{\mA}{\mathbf{A}}
\newcommand{\mU}{\mathbf{U}}
\newcommand{\mV}{\mathbf{V}}
\newcommand{\mC}{\mathbf{C}}

The global method presented in Sec.\,\ref{sec:global} involves the computation of the inverse matrix $\left(\left(\mS_\con\right)^{-1}-\mS_{\iC\iC}\right)^{-1}$, which represents the most costly part of the computation of the scattering matrix of a connected supersystem. 
Many applications in quantitative parametric analysis or surrogate optimization of reconfigurable systems \cite{tapie2023systematic} require repeated evaluations of the same connected supersystem that only differ regarding details of one or a few subsystem(s).
To alleviate the computational cost of such repeated evaluations of the same connection scheme, we develop a technique to perform low-rank updates of the inverted matrix in the present section. 

This so-called updatability technique can also be used after having applied the reduction methods discussed in Sec.~\ref{sec:reducibility}. If the connection system has no free ports after reduction, the mathematical form remains that of cascade-loading as in the global method. If the connection system involves free ports after reduction, the Redheffer star product is required, as detailed in Sec.~\ref{sec:reducibility}. In that case, two matrix inversions are involved (see Eq.~(\ref{eq:redheffer_aux})). These can also be updated using the Woodbury matrix identity, but we do not explicitly describe this case in the following. Instead, for the sake of conciseness, we limit our description to the case of problems formulated in the form of cascade loading.

\subsection{Update method using Woodbury matrix identity}

We consider the case in which an arbitrary supersystem $\Gl$ comprising $n$ subsystems $\E_i$ has been defined; its scattering matrix is given by
\begin{equation}
    \mS^\Gl = \mathrm{blockdiag}\left\{\mS^{\E_1},\dots,\mS^{\E_n}\right\}.
\end{equation}
We further consider that a connection scheme of the supersystem has been fixed; the scattering matrix of the connection system is denoted by $\mS^\Gl_\con$. The connection scheme could involve only $\delta$-connections in the global method, or some subsystems (other than those already included in $\Gl$) as a result of the reductions proposed in Sec.~\ref{sec:reducibility}. In the latter case, as already stated, we limit our analysis to cases in which the connection system has no free ports such that the connected supersystem can be evaluated in terms of cascade loading (rather than requiring the Redheffer star product). Thus, following Eq.\,(\ref{eq:generic}), the scattering matrix $\tilde{\mS}^\Gl$ resulting from the considered connection scheme for this supersystem is given by
\begin{equation}
    \tilde{\mS}^\Gl = \mS^\Gl_{\iN\iN} + \mS^\Gl_{\iN\iC} \, \mSi \, \mS^\Gl_{\iC\iN},
\end{equation}
where we introduced
\begin{equation}
    \mSi = \left( \left({\mS^\Gl_\con}\right)^{-1} - \mS^\Gl_{\iC\iC} \right)^{-1}
\end{equation}
 for notational convenience.

Let us now consider that the scattering matrix of the $j$th subsystem, $\mathrm{E}_j$, has changed. Because we assume that the connection system remains fixed, $\mathrm{E}_j$ must be included in the supersystem (which is automatically the case in the global method but not necessarily upon reduction). Moreover, the number of ports of $\mathrm{E}_j$ must hence remain the same. We denote the updated version of $\mS^\Gl$ by ${\mS^\Gl}^\prime$; its constituent blocks ${\mS^\Gl_{\iN\iN}}^\prime$, ${\mS^\Gl_{\iN\iC}}^\prime$, ${\mS^\Gl_{\iC\iN}}^\prime$ and  ${\mS^\Gl_{\iC\iC}}^\prime$ are obtained straightforwardly without significant computational cost. However, evaluating the updated scattering matrix $\tilde{\mS}^{\Gl\prime}$ of the connected system is not straightforward because it is given by
\begin{equation}
    \tilde{\mS}^{\Gl\prime} = {\mS^\Gl_{\iN\iN}}^\prime + {\mS^\Gl_{\iN\iC}}^\prime
    \, \mSi^\prime \, {\mS^\Gl_{\iC\iN}}^\prime,
\label{eq:updatedSmatrix}
\end{equation}
where
\begin{equation}
    \mSi^\prime = \left( \left({\mS^\Gl_\con}\right)^{-1} - {\mS^\Gl_{\iC\iC}}^\prime \right)^{-1}
\end{equation}
involves a matrix inversion.  The interest of the update method discussed in this section lies in efficiently obtaining the updated inverse $\mSi^\prime$.

Instead of evaluating $\mSi^\prime$ from scratch, we seek to use the Woodbury matrix identity to obtain $\mSi^\prime$ based on our knowledge of $\mSi$ and the change from $\mS^\Gl_{\iC\iC}$ to ${\mS^\Gl_{\iC\iC}}^\prime$ that led to this update. Let us define $\mathbf{A} = \left({\mS^\Gl_\con}\right)^{-1} - {\mS^\Gl_{\iC\iC}}^\prime$ such that we are interested in $\mSi^\prime = \mathbf{A}^{-1}$. The change  $\mS^\Gl_{\iC\iC} \rightarrow {\mS^\Gl_{\iC\iC}}^\prime$ implies $\mathbf{A} \rightarrow \mathbf{A} + \Delta\mathbf{A}$, where $ \Delta\mathbf{A} = {\mS^\Gl_{\iC\iC}}^\prime-{\mS^\Gl_{\iC\iC}}$. Provided that we can express $\Delta\mathbf{A}$ as the product of three matrices, i.e., $\Delta\mathbf{A} = \mathbf{UCV}$, we will be able to apply the Woodbury matrix identity~\cite{hager1989updating}:
\begin{equation}
\!\! {\left(\mA \!\!+\!\! \mU\mC\mV\right)}^{-1} \!\!=\!\! \mA^{-1} \!\!-\! \mA^{-1}\mU {\left(\mC^{-1} \!\!+\!\! \mV\mA^{-1}\mU\right)}^{-1} \mV\mA^{-1}.
\label{eq_woodbury}
\end{equation}
We can meet this goal using the following definitions of $\mU$, $\mC$ and $\mV$:
\begin{equation}
\mC = -\Delta \mS^\Gl_{\iC_j\iC_j} = -\Delta \mS^{\E_j}_{\{\iC\iC\}}
\end{equation}
\begin{equation}
\setlength\arraycolsep{1pt}
\mV = \ \begin{bNiceMatrix}[first-col, first-row]
        & \iC_1 & \dots & \iC_j & \dots & \iC_n \\
\iC_j   & \mO   & \dots & \mI   & \dots & \mO
\end{bNiceMatrix}, \quad
\mU = \mV^\mathrm{T},
\end{equation}
where $\iC_j \subseteq \iC$ are the connected ports of the updated subsystem $\mathrm{E}_j$. The rank of the update is hence $n(\iC_j)$.

Noting that the products between $\mSi$, $\mU$ and $\mV$ simplify as
\begin{equation}
\mSi\mU=\mSi_{\iC\iC_j}, \quad \mV\mSi=\mSi_{\iC_j\iC}, \quad \mV\mSi\mU=\mSi_{\iC_j\iC_j},
\end{equation}
and applying the Woodbury identity from Eq.~(\ref{eq_woodbury}), we obtain 
\begin{equation}
    \mSi^\prime = \mSi -  \mSi_{\iC\iC_j} \left(\left({-\Delta \mS^{\E_j}_{\{\iC\iC\}}}\right)^{-1} + \mSi_{\iC_j\iC_j} \right)^{-1} \mSi_{\iC_j\iC}.
\label{eq:updatedinversematrix}
\end{equation}
In using Eq.\,(\ref{eq:updatedinversematrix}) to obtain $\mSi^\prime$, we perform two inversions of $n\left(\iC_j\right)\times n\left(\iC_j\right)$ matrices and two products involving matrices of sizes $n\left(\iC_j\right)\times n\left(\iC_j\right)$ and $n\left(\iC_j\right) \times n\left(\iC\right)$, instead of inverting one $n\left(\iC\right)\times n\left(\iC\right)$ matrix. Applying Eq.\,(\ref{eq:updatedSmatrix}) subsequently provides the updated resulting scattering matrix of the connected supersystem.

\begin{figure}[h]
    \centering
    \includegraphics[width=0.99\columnwidth]{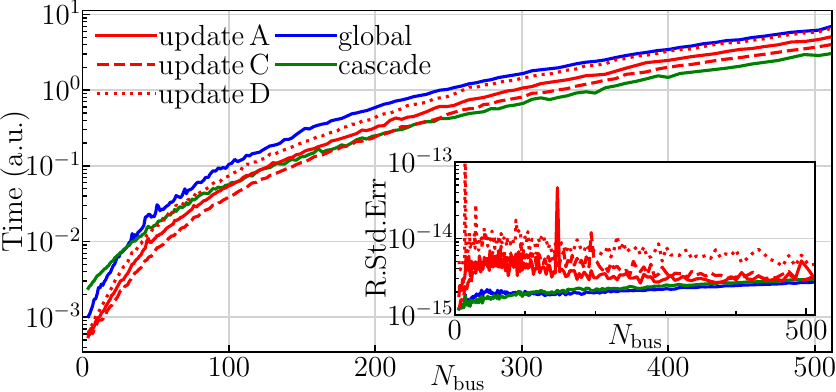}
    \caption{Computation time (main figure) and relative standard error (inset) for the evaluation of the scattering matrix of the meta-network of Fig.\,\ref{fig:meta-network} after one of its constituent subsystems has been updated, using three methods: global (Sec.~\ref{sec:global}), cascade (iterative application of the Redheffer star product, adding on one subsystem after the other) and update (this section). Similar to Sec.~\ref{sec:reducibility}, all evaluations are performed analytically using graphs as scattering systems (see Appendix~\ref{Appendix_graphs} for details). $N_\mathrm{bus}$ is again the number of ports via which any two subsystems are connected. Therefore, the update of the system $\A$, $\C$ or $\D$ implies a change of a subsystem with $3N_\mathrm{bus}$, $2N_\mathrm{bus}$ or $4N_\mathrm{bus}$ ports, respectively. (The update of $\B$ yields similar results as the update of $\A$.) The total number of ports is $12N_\mathrm{bus}$.}
    \label{fig:update_benchmark}
\end{figure}

As presented in Fig.\,\ref{fig:update_benchmark}, the update method can provide significant speedups with a negligible loss of accuracy (relative standard error $\lesssim 10^{-14}$, see inset). The speedup depends on the ratio $n\left(\iC_j\right)/n\left(\iC\right)$ and the overall number of ports. The former is evidenced by the comparison of updating A, C or D in Fig.\,\ref{fig:update_benchmark} for which the ratio $n\left(\iC_j\right)/n\left(\iC\right)$ is 3/12, 2/12 or 4/12, respectively. There is a clear hierarchy between the three red curves, where the largest speedup corresponds to the case with the smallest value of $n\left(\iC_j\right)/n\left(\iC\right)$, i.e., in the case of updating C. This is in line with related findings in the context of RIS-parametrized radio environment for updating a channel upon a change of the RIS configuration using the Woodbury identity (see Fig.~2 in~\cite{prod2023efficient}). The dependence on $N_\mathrm{bus}$ is  once again related to the memory access latency, similar to Fig.~\ref{fig:reducibility_benchmark}. For very large values of $N_\mathrm{bus}$ (note that for $N_\mathrm{bus}=500$ we are considering a total of $6\times 10^4$ ports), memory access becomes the bottleneck and thus the cascade approach is the fastest in these cases.

\section{Closed-form recovery of power waves passing through connected ports}
\label{sec:generalizedconnect_potentials}

The global method presented in Sec.\,\ref{sec:global} not only establishes a transparent closed-form approach to evaluate the scattering properties of an arbitrarily complex connection of systems but it also enables the non-iterative recovery of the power waves passing through the connected ports. Recovering this information has previously been explored in dedicated studies based on iterative methods~\cite{belenguer_krylovs_2013,diaz_caballero_extending_2014} motivated by the fact that this information allows for the assessment of device vulnerabilities (such as those arising from multipactor effects or corona discharges in components), and the reconstruction of the field distribution inside the connected system given sufficient a priori knowledge about its geometry.
Our power wave recovery method can be applied to reduced supersystems as presented in Sec.\,\ref{sec:reducibility} and the updatability method of Sec.\,\ref{sec:updatability} can be used to speed up the recovery after an update of subsystems. In the following, we consider a generic cascade-loading problem involving a system $\mS^\meta$ with some free and some connected ports and a connection system $\mS_\con$ without any free ports. Applied to the meta-network, $\mS^\mathrm{ABCD}$ would be $\mS^\meta$ and $\mS^\mathrm{ABCD}_\con$ would be $\mS_\con$.

A direct by-product of the derivation of cascade loading with scattering parameters in Appendix\,\ref{Appendix_Derivations_S} is the insight that the power wave $\mathbf{b}_\mathcal{C}\in\mathbb{C}^{n(\mathcal{C}) \times 1}$ exiting the connected ports of the supersystem upon injecting the power wave $\mathbf{a}_\mathcal{N}\in\mathbb{C}^{n(\mathcal{N}) \times 1}$ into the supersystem's free ports is 
\begin{equation}
    \mathbf{b}_\mathcal{C} = \left({\mS}_\mathrm{con}^{-1}-\mS^\meta_{\iC\iC}\right)^{-1}
    \mS^\meta_{\iC\iN} \mathbf{a}_\iN.
\label{eq:Scascade-bC_repeat}
\end{equation}
Moreover, using $\mathbf{b}_\mathcal{C} = {\mS}_\mathrm{con} \mathbf{a}_\mathcal{C}$ in analogy with Eq.\,(\ref{eq:Scascade-S2}) from Appendix~\ref{Appendix_Derivations_S}, where $\mathbf{a}_\mathcal{C}\in\mathbb{C}^{n(\mathcal{C}) \times 1}$ is the power wave entering the connected ports of the supersystem, we can relate  $\mathbf{a}_\mathcal{C} $ to $\mathbf{a}_\mathcal{N}$ with
\begin{equation}
    \mathbf{a}_\mathcal{C} = {\mS}_\mathrm{con}^{-1} \mathbf{b}_\mathcal{C}  = {\mS}_\mathrm{con}^{-1} \left({\mS}_\mathrm{con}^{-1}-\mS^\meta_{\iC\iC}\right)^{-1}
    \mS^\meta_{\iC\iN} \mathbf{a}_\iN.
\label{eq:Scascade-bC_repeat}
\end{equation}

It is usually more relevant to compute the power wave potentials $\boldsymbol{\psi}_\mathcal{C} = \mathbf{a}_\mathcal{C} +\mathbf{b}_\mathcal{C} $ at the connected ports and the power wave fluxes $\boldsymbol{\phi}_\mathcal{C} = \mathbf{a}_\mathcal{C} -\mathbf{b}_\mathcal{C} $ through the connected ports because these are proportional to voltages and currents, respectively (see Appendix~\ref{Appendix_Definitions}). The incoming power wave $\mathbf{a}_\iN$ can be directly mapped to these quantities of interest via $\boldsymbol{\psi}_\mathcal{C} =  \mathbf{\Psi}^\meta_{\iC\iN} \mathbf{a}_\iN$ and $\boldsymbol{\phi}_\mathcal{C} =  \mathbf{\Phi}^\meta_{\iC\iN} \mathbf{a}_\iN$, where
\begin{equation}
    \mathbf{\Psi}^\meta_{\iC\iN} = \left( \mS_\con^{-1} + \mI \right)
    \left( \mS_\con^{-1} - \mS^\meta_{\iC\iC} \right)^{-1} \mS^\meta_{\iC\iN},
\end{equation}
\begin{equation}
    \mathbf{\Phi}^\meta_{\iC\iN} = \left( \mS_\con^{-1} - \mI \right)
    \left( \mS_\con^{-1} - \mS^\meta_{\iC\iC} \right)^{-1} \mS^\meta_{\iC\iN}.
\end{equation}

\begin{rem}
Let $\iC_i \in \mathcal{C}$ and $\iC_j\in \mathcal{C}$ be two sets of ports connected by a set of $\delta$-connections. In this case, the ports $\iC_i$ and $\iC_j$ are inherently merged together, yet they have mutually opposite orientations, implying
\begin{equation}
\mathbf{\Psi}^\meta_{\iC_i\iN}=\mathbf{\Psi}^\meta_{\iC_j\iN}, \quad \mathbf{\Phi}^\meta_{\iC_i\iN}=-\mathbf{\Phi}^\meta_{\iC_j\iN}.
\end{equation}
\end{rem}

\section{Transposition to Impedance and \\Admittance Parameters}
\label{sec_impedance_admittance}

So far, our analysis was limited to representations of the involved systems in terms of scattering parameters. In this section, we extend our results to impedance and admittance parameters which are alternative equivalent ways of characterizing multi-port networks.

\subsection{Impedance parameters}

For the analysis of our meta-network in terms of impedance parameters, we start by defining the impedance matrix $\mZ^\meta$ of the supersystem in a similar way as for scattering parameters in Eq.\,(\ref{eq:meta-network_S_super}):
\begin{equation}
    \mZ^\super = \mathrm{blockdiag}\left\{ \mZ^\A,\mZ^\B,\mZ^\C,\mZ^\D \right\},
\end{equation}
where $\mZ^\X$ denotes the impedance matrix of the subsystem $\X$. The conversion between scattering parameters and impedance parameters is well established and provided in Appendix~\ref{Appendix_Definitions} for reference.

Obtaining $\mZ_\con$ can be challenging if the connection system involves $\delta$-connections. Indeed, the impedance of a delayless, lossless, reflectionless, reciprocal two-port system (i.e., a $\delta$-connection) cannot be defined properly, and by extension the impedance matrix of a connection system based on a set of $\delta$-connections cannot be defined properly. This issue manifests itself in the conversion from $\mS_\con$ to $\mZ_\con$ which requires the inversion of $\left(\mI-\mS_\con\right)$ but this matrix is not invertible in the case of a connection system involving $\delta$-connections. A work-around consists in multiplying the unity values within $\mS_\con$ by $\left(1-\epsilon\right)$, with $\epsilon\to0$, before evaluating $\left(\mI-\mS_\con\right)^{-1}$ as part of the conversion to impedance parameters. This will yield the impedance matrix $\mZ_\con^\epsilon$ of a connection system in which ideal $\delta$-connections are replaced by \textit{quasi}-$\delta$-connections.

The impedance matrix $\tilde{\mZ}^\super$ of the connected supersystem is then given by
\begin{equation}
    \tilde{\mZ}^{\super} = \mZ^{\super}_{\iN\iN} - \mZ^{\super}_{\iN\iC}
    \left( \mZ_\con^{\super,\epsilon} + \mZ^{\super}_{\iC\iC} \right)^{-1}
    \mZ^{\super}_{\iC\iN},
\label{eq:meta-network-cascade-global-Z}
\end{equation}
which is the impedance-parameter analogue of Eq.~(\ref{eq:generic}). A derivation is provided for completeness in Appendix~\ref{Appendix_Derivations_Z}.
It is important to note that $\mZ_\con^{\super,\epsilon}$ specializes to $\mZ_\con^\super$ only if the meta-network was fully reduced (using the reducibility established in Sec.~\ref{sec:reducibility}) such that the connection system does not involve any $\delta$-connections that must be replaced by quasi-$\delta$-connections when working with impedance parameters.

\begin{figure}[h]
    \centering
    \includegraphics[width=1\columnwidth]{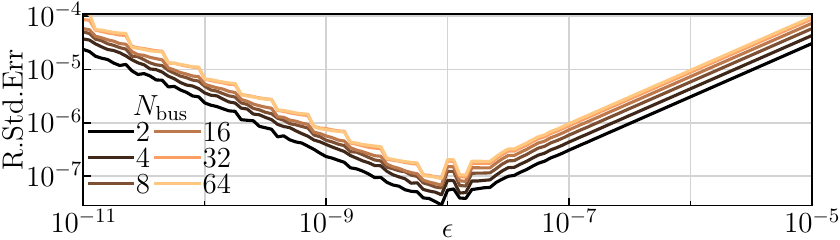}
    \caption{Relative standard error $\mathrm{std}\left(\tilde{\mZ}^\meta-\mZ^\metaref\right)/\left\langle\left|\mZ^\metaref\right|\right\rangle$ of the impedance matrix $\tilde{\mZ}^\mathrm{ABCD}$ of the connected supersystem as a function of the value of $\epsilon$. $\mZ^\metaref$ is obtain by evaluating the scattering parameters of a given supergraph and converting them to impedance parameters; meanwhile, $\tilde{\mZ}^\meta$ is obtained by evaluating the scattering parameters of the individual subgraphs, converting them to impedance parameters, and evaluating their connection using Eq.~(\ref{eq:meta-network-cascade-global-Z}). $N_\mathrm{bus}$ is the number of ports via which subgraphs are connected to each other.}
    \label{fig:metaZ_accuracy}
\end{figure}

The requirement for a replacement of $\delta$-connections with quasi-$\delta$-connections that arises when working with impedance parameters naturally brings about a loss of accuracy when dealing with not-fully-reducible connection schemes. Considering the meta-network example (which is not fully reducible, as explained in Sec.~\ref{sec:reducibility}), we present in Fig.\,\ref{fig:metaZ_accuracy} the relative standard error of the obtained impedance matrix $\tilde{\mZ}^\mathrm{ABCD}$ with respect to the ground-truth reference one $\mZ^\metaref$, as a function of the value of $\epsilon$. The reference $\mZ^\metaref$ is obtained by ``gluing together'' the graphs that generated the scattering matrices of the constituent subsystems, generating the corresponding combined scattering matrix $\mS^\metaref$, and converting it to impedance parameters. As the value of $\epsilon$ decreases, the relative standard error initially decreases (i.e., the accuracy improves), until a critical value of $\epsilon$ is reached at $\epsilon\sim10^{-8}$. At this sweet spot, the relative standard error is on the order of $10^{-7}$, subtly depending on the size of the system. Beyond this critical value, the relative standard error increases again because $\epsilon$ is too small in this regime and cannot prevent the quasi-singular nature of the matrix to be inverted. However, even using the ideal value of $\epsilon$, the relative standard error is orders of magnitude larger than in previous sections based on scattering parameters. Therefore, we conclude that there is a fundamental benefit in working with scattering rather than impedance parameters when considering connection schemes that are not fully reducible.

The derivation of Eq.~(\ref{eq:generic}) in Appendix~\ref{Appendix_Derivations_Z} also directly yields the equations required to retrieve the voltages at the connected ports, see Eq.~(\ref{eq_volt_z_internal}), and the currents flowing through the connected ports, see Eq.~(\ref{eq:Zcascade-JC}), in terms of impedance parameters. 

For the sake of completeness, we further mention that the impedance matrix of the connection of two systems with free ports, as presented in Fig.\,\ref{fig:genericcascade}, can be obtained with an impedance-parameter analogue of the Redheffer star product~\cite{jabotinski2018efficient}:
\begin{multline}
\tilde{\mZ}^{\U\V}_{\iN\iN} = \\
\setlength\arraycolsep{0pt}
\!\!\! \mZ^{\U\V}_{\iN\iN} -\mZ^{\U\V}_{\iN\iC} \left(
\left(\mZ^{\U}_{\iC_\U\iC_\U} + \mZ^{\V}_{\iC_\V\iC_\V}\right)^{-1} \! \otimes \!
\begin{bmatrix} 1 & -1 \\ -1 & 1 \end{bmatrix} 
\right) \mZ^{\U\V}_{\iC\iN}.
\end{multline}

\subsection{Admittance parameters}
\newcommand{\mY}{\ensuremath{\mathbf{Y}}}

The use of admittance parameters suffers from the same issue regarding the description of $\delta$-connections as the use of impedance parameters. Definitions of admittance parameters and their relation to impedance and scattering parameters are provided in Sec.~\ref{Appendix_Definitions}. A derivation of cascade loading with admittance parameters is provided for completeness in Appendix~\ref{sec_derivation_admittance} and yields 
\begin{equation}
    \tilde{\mY}^\meta = \mY^\meta_{\iN\iN} + \mY^\meta_{\iN\iC}
    \left( \mY_\con^\epsilon + \mY^\meta_{\iC\iC} \right)^{-1}
    \mY^\meta_{\iC\iN},
    \label{cascade_Y_main}
\end{equation}
where $\mY_\con^\epsilon$ may again require the consideration of quasi-$\delta$-connections instead of ideal $\delta$-connections, unless the meta-network has been fully reduced. The derivation of Eq.~(\ref{cascade_Y_main}) in Appendix~\ref{sec_derivation_admittance} also directly yields the equations required to retrieve the voltages at the connected ports, see Eq.~(\ref{voltage_connected_ports_admittance}), and the currents flowing through the connected ports, see Eq.~(\ref{current_connected_ports_admittance}), in terms of admittance parameters.

\section{Conclusion}
\label{sec_conclusion}

To summarize, we have illustrated an updatable closed-form technique for the evaluation of arbitrarily complex connections between multi-port systems. Thereby, repeated evaluations of the same connection scheme for modified versions of a constituent subsystem can be realized by updating the outcome of a previous evaluation of the connection scheme as opposed to starting from scratch. 
Along the way, we have presented unified equivalence principles for the interpretation of multi-port system connections in a pedagogical manner, we have established reducibility methods for the closed-form global technique, and we have identified a closed-form derivation of the power waves flowing through connected ports. We have rigorously validated our results with physics-compliant studies of graphs and conducted exhaustive statistical analyses of computational efficiency benefits enabled by reducibility and updatability. Moreover, we established a fundamental benefit of working with scattering parameters for connection schemes that are not fully reducible.

Looking forward, we envision to leverage the techniques developed in this paper for the design of large composite wave systems as well as the characterization of large \textit{reconfigurable} composite wave systems.

\appendices

\section{Definitions}
\label{Appendix_Definitions}

\newcommand{\ma}{\ensuremath{\mathbf{a}}}
\newcommand{\mb}{\ensuremath{\mathbf{b}}}
\newcommand{\mJ}{\ensuremath{\mathbf{J}}}

Let $V_i$ denote the voltage across the two terminals of the $i$th port of an $N$-port system, and let $J_i$ denote the inward current flowing into the system via its $i$th port. The impedance matrix $\mathbf{Z}\in\mathbb{C}^{N\times N}$ relates the voltages $\mathbf{V}\in\mathbb{C}^{N\times 1}$ and inward currents $\mathbf{J}\in\mathbb{C}^{N\times 1}$ via
\begin{equation}
    \mV = \mZ \mJ.          \label{eq:def-Z}
\end{equation}
This relation can equivalently be expressed in terms of the admittance matrix $\mathbf{Y}\in\mathbb{C}^{N\times N}$ as 
\begin{equation}
    \mJ = \mY \mV          \label{eq:def-Y}
\end{equation}
because the admittance matrix is defined as the inverse of the impedance matrix
\begin{equation}
    \mathbf{Y} = \mathbf{Z}^{-1}.
\end{equation}

The power wave formalism~\cite{kurokawa_power_1965} reformulates the problem by relating the input power wave $\mathbf{a}\in \mathbb{C}^{N \times 1}$ to the output power wave $\mathbf{b}\in \mathbb{C}^{N \times 1}$ via the scattering matrix  $\mS \in \mathbb{C}^{N\times N}$ following
\begin{equation}
    \mb = \mS \ma.          \label{eq:def-S}
\end{equation}
The $i$th component of the input and output power waves are defined as
\begin{equation}
    a_i = \frac{V_i+Z_i J_i}{2\sqrt{\mathrm{Re}(\left|Z_i\right|)}},        \quad
    b_i = \frac{V_i - Z_i^\ast J_i}{2\sqrt{\mathrm{Re}(\left|Z_i\right|)}},
\label{eq:def-powerwaves}
\end{equation}
where $Z_i$ is the characteristic impedance of the transmission line attached to the $i$th port.

Distinguishing the terms related to the voltages and the currents within the power waves expression, we introduce the ``power wave potential'' ${\psi}_i$ and the ``power wave flux'' ${\phi}_i$ as
\begin{equation}
    {\psi}_i = {a}_i\!+\!{b}_i = \frac{1}{\sqrt{\mathrm{Re}(\left|Z_i\right|)}} V_i, \quad
    {\phi}_i = {a}_i\!-\!{b}_i = \frac{Z_i}{\sqrt{\mathrm{Re}(\left|Z_i\right|)}} J_i.
    \label{def_eq_psi_phi}
\end{equation}
Analogous to Kirchhoff's voltage and current laws, the power wave potentials $\boldsymbol{\psi} \in \mathbb{C}^{N\times 1}$ and power wave fluxes $\boldsymbol{\phi} \in \mathbb{C}^{N\times 1}$ are respectively subject to the constraints of continuity and conservation. They are also directly related to the input and output power waves by
\begin{equation}
    \boldsymbol{\psi} = \left(\mI+\mS\right)\mathbf{a}, \quad \boldsymbol{\phi} = \left(\mI-\mS\right)\mathbf{a}.
    \label{eq_Psi_I_S_a}
\end{equation}

The relation between scattering matrix and impedance matrix is given by~\cite{kurokawa_power_1965}
\begin{equation}
    \mathbf{S} = \mathbf{F}(\mathbf{Z}-\mathbf{G}^\dagger)(\mathbf{Z}+\mathbf{G})^{-1}\mathbf{F}^{-1},
    \label{eq_S2Zgeneral}
\end{equation}
where $\mathbf{F}$ and $\mathbf{G}$ are diagonal matrices whose $i$th diagonal entries are $F_{ii} = \frac{1}{2\sqrt{|\mathrm{Re}(Z_i)|}}$ and $G_{ii} = Z_i$. In most typical cases, the characteristic impedances of all the transmission lines are constrained to the same real value $Z_0 \in \mathbb{R}$ such that the expression from Eq.\,(\ref{eq_S2Zgeneral}) simplifies to the commutative product
\begin{equation}
    \mathbf{S} = (\mathbf{Z}-Z_0\mathbf{I})(\mathbf{Z}+Z_0\mathbf{I})^{-1}= (\mathbf{Z}+Z_0\mathbf{I})^{-1}(\mathbf{Z}-Z_0\mathbf{I}).
    \label{eqSZallZ0}
\end{equation}

\section{Derivations}
\label{Appendix_Derivations}

In this appendix, we consider the basic cascade setup displayed in Fig.\,\ref{fig:innnerouterequivalency}. We refer to the top system by (1) and to the bottom system by (2). The sets $\iN$ and $\iC$ include the indices of the free ports and connected ports of (1), respectively; meanwhile, (2) has no free ports. Each of the ports of (1) whose index is included in $\mathcal{C}$ is connected via a $\delta$-connection to a different port of (2).

\subsection{Impedance parameters}
\label{Appendix_Derivations_Z}

The impedance  matrices characterizing (1) and (2) are $\mathbf{Z}^{(1)}\in\mathbb{C}^{n(\mathcal{C}\cup\mathcal{N})\times n(\mathcal{C}\cup\mathcal{N})}$ and $\mathbf{Z}^{(2)}\in\mathbb{C}^{n(\mathcal{C})\times n(\mathcal{C})}$, respectively. According to Eq.\,(\ref{eq:def-Z}), the following relations hold:
\begin{subequations}
\begin{align}
        \mV^{(1)} & = \mZ^{(1)} \mJ^{(1)},                                                     \label{eq:Zcascade-Z1} \\
    \mV^{(2)} & = \mZ^{(2)} \mJ^{(2)}.                                                     \label{eq:Zcascade-Z2}
\end{align}
\end{subequations}
By partitioning $\mathbf{Z}^{(1)}$ into a $2\times 2$ block matrix and the vectors  $\mathbf{V}^{(1)}$ and  $\mathbf{J}^{(1)}$ into corresponding $2 \times 1$ block vectors, we can cast Eq.\,(\ref{eq:Zcascade-Z1}) into the following form:
\begin{subequations}
\begin{align}
    \mV^{(1)}_\iN & = \mZ^{(1)}_{\iN\iN} \mJ_\iN^{(1)} + \mZ^{(1)}_{\iN\iC} \mJ_\iC^{(1)}, \label{eq:Zcascade-Z1N} \\
    \mV^{(1)}_\iC & = \mZ^{(1)}_{\iC\iN} \mJ_\iN^{(1)} + \mZ^{(1)}_{\iC\iC} \mJ_\iC^{(1)}. \label{eq:Zcascade-Z1C}
\end{align}
\end{subequations}
At each pair of connected ports, the voltage across the two terminals of each port must be equal; moreover, the inward current of one port must equal the negative of the inward current of the other port:
\begin{subequations}
\begin{align}
    \mV^{(1)}_\iC & = \mV^{(2)},    \label{eq:Zcascade-eqV} \\
    \mJ^{(1)}_\iC & = -\mJ^{(2)}.   \label{eq:Zcascade-eqJ}
\end{align}
\end{subequations}
Inserting Eqs.~(\ref{eq:Zcascade-eqV}, \ref{eq:Zcascade-eqJ}) into Eq.\,(\ref{eq:Zcascade-Z2}) yields
\begin{equation}
    \mV^{(1)}_\iC = -\mZ^{(2)} \mJ^{(1)}_\iC. \label{eq:Zcascade-VC}
\end{equation}
Inserting Eq.\,(\ref{eq:Zcascade-VC}) into Eq.\,(\ref{eq:Zcascade-Z1C}) and rearranging yields
\begin{equation}
    \mJ^{(1)}_\iC = - {\left( \mZ^{(1)}_{\iC\iC} + \mZ^{(2)} \right)}^{-1} \mZ^{(1)}_{\iC\iN} \mJ^{(1)}_\iN. \label{eq:Zcascade-JC}
\end{equation}
Finally, inserting Eq.\,(\ref{eq:Zcascade-JC}) into Eq.\,(\ref{eq:Zcascade-Z1N}) yields
\begin{equation}
    \mV^{(1)}_\iN = \mZ^{(12)} \mJ^{(1)}_\iN,
    \label{free_port_imp_rel}
\end{equation}
where $\mZ^{(12)}$ is the impedance matrix of the cascaded system:
\begin{equation}
    \mZ^{(12)} = \mZ^{(1)}_{\iN\iN} - \mZ^{(1)}_{\iN\iC} {\left( \mZ^{(1)}_{\iC\iC} + \mZ^{(2)} \right)}^{-1} \mZ^{(1)}_{\iC\iN}. \label{eq:Zcascade-final}
\end{equation}

The currents $\mathbf{J}^{(1)}_\iC = -\mathbf{J}^{(2)}$ flowing through the connected ports are directly given by Eq.\,(\ref{eq:Zcascade-JC}) as a function of the currents $\mathbf{J}^{(1)}_\iN$ flowing into the free ports of (1). The voltages $\mathbf{V}^{(1)}_\iC = \mathbf{V}^{(2)}$ across the terminals of the connected ports are given by Eq.\,(\ref{eq:Zcascade-VC}) in terms of $\mathbf{J}^{(1)}_\iC$ and can hence be expressed in terms of the currents $\mathbf{J}^{(1)}_\iN$ flowing into the free ports of (1) as
\begin{equation}
    \mV^{(1)}_\iC = \mV^{(2)} = \mZ^{(2)} {\left( \mZ^{(1)}_{\iC\iC} + \mZ^{(2)} \right)}^{-1} \mZ^{(1)}_{\iC\iN} \mJ^{(1)}_\iN.
    \label{eq_volt_z_internal}
\end{equation}

\subsection{Admittance parameters}
\label{sec_derivation_admittance}

The admittance  matrices characterizing (1) and (2) are $\mathbf{Y}^{(1)}\in\mathbb{C}^{n(\mathcal{C}\cup\mathcal{N})\times n(\mathcal{C}\cup\mathcal{N})}$ and $\mathbf{Y}^{(2)}\in\mathbb{C}^{n(\mathcal{C})\times n(\mathcal{C})}$, respectively. 

Following a procedure analogous to that outlined in Appendix~\ref{Appendix_Derivations_Z}, we find that the admittance matrix $\mathbf{Y}^{(12)}$ of the cascaded system is
\begin{equation}
    \mY^{(12)} = \mY^{(1)}_{\iN\iN} - \mY^{(1)}_{\iN\iC} {\left( \mY^{(1)}_{\iC\iC} + \mY^{(2)} \right)}^{-1} \mY^{(1)}_{\iC\iN}. \label{eq:Zcascade-final}
\end{equation}
At the free ports $\iN$, the inward currents $\mathbf{J}^{(1)}_\iN$ are related to the voltages $\mathbf{V}^{(1)}_\iN$ through $\mY^{(12)}$ according to
\begin{equation}
    \mathbf{J}^{(1)}_\iN = \mY^{(12)} \mathbf{V}^{(1)}_\iN.
\end{equation}

In analogy with Eq.\,(\ref{eq:Zcascade-JC}), the voltages across the terminals of the connected ports are given in terms of the voltages across the terminals of the free ports of (1) by
\begin{equation}
    \mathbf{V}^{(1)}_\iC = \mathbf{V}^{(2)} = - {\left( \mY^{(1)}_{\iC\iC} + \mY^{(2)} \right)}^{-1} \mY^{(1)}_{\iC\iN} \mV^{(1)}_\iN.
    \label{voltage_connected_ports_admittance}
\end{equation}

In analogy with Eq.\,(\ref{eq_volt_z_internal}), the currents $\mathbf{J}^{(1)}_\iC = -\mathbf{J}^{(2)} $ flowing through the connected ports of (1) are given in terms of the voltages $ \mathbf{V}^{(1)}_\iN$ across the terminals of the free ports of (1) by
\begin{equation}
        \mJ^{(1)}_\iC = -\mJ^{(2)} = \mY^{(2)} {\left( \mY^{(1)}_{\iC\iC} + \mY^{(2)} \right)}^{-1} \mY^{(1)}_{\iC\iN} \mV^{(1)}_\iN.
    \label{current_connected_ports_admittance}
\end{equation}

\subsection{Scattering parameters}
\label{Appendix_Derivations_S}

The scattering matrices characterizing (1) and (2) are $\mathbf{S}^{(1)}\in\mathbb{C}^{n(\mathcal{C}\cup\mathcal{N})\times n(\mathcal{C}\cup\mathcal{N})}$ and $\mathbf{S}^{(2)}\in\mathbb{C}^{n(\mathcal{C})\times n(\mathcal{C})}$, respectively.  According to Eq.\,(\ref{eq:def-S}), the following power wave relations hold:
\begin{subequations}
    \begin{align}
    \mb^{(1)} & = \mS^{(1)} \ma^{(1)}.                                                     \label{eq:Scascade-S1}\\
    \mb^{(2)} & = \mS^{(2)} \ma^{(2)}.                                                     \label{eq:Scascade-S2}       
    \end{align}
\end{subequations}
By partitioning $\mathbf{S}^{(1)}$ into a $2\times 2$ block matrix as in Eq.\,(\ref{eq_block}) and by partitioning the power wave vectors $\mathbf{a}^{(1)}$ and $\mathbf{b}^{(1)}$ into corresponding $2\times 1$ block vectors, Eq.\,(\ref{eq:Scascade-S1}) can be recast as
\begin{subequations}
\begin{align}
    \mb^{(1)}_\iN & = \mS^{(1)}_{\iN\iN} \ma^{(1)}_\iN + \mS^{(1)}_{\iN\iC} \ma^{(1)}_\iC. \label{eq:Scascade-S1N} \\
    \mb^{(1)}_\iC & = \mS^{(1)}_{\iC\iN} \ma^{(1)}_\iN + \mS^{(1)}_{\iC\iC} \ma^{(1)}_\iC. \label{eq:Scascade-S1C}
\end{align}
\end{subequations}
At the connected ports, the power waves going into one system must be equal to the outgoing power waves of the other system:
\begin{subequations}
\begin{align}
    \mathbf{a}^{(1)}_\iC=\mathbf{b}^{(2)}, \label{eq:Scascade-eqa1} \\
    \mathbf{a}^{(2)}=\mathbf{b}^{(1)}_\iC. \label{eq:Scascade-eqa2}
\end{align}
\end{subequations}
Inserting Eq.\,(\ref{eq:Scascade-eqa2}) into Eq.\,(\ref{eq:Scascade-S2}), and Eq.\,(\ref{eq:Scascade-eqa1}) into Eq.\,(\ref{eq:Scascade-S1C}), combining the two equations, and re-arranging yields the following expression for the power wave $\mb^{(2)}$ exiting (2) upon injecting the power wave $\ma_\iN^{(1)}$ into the free ports of (1):
\begin{equation}
    \mathbf{b}^{(2)} = \left({\mS^{(2)}}^{-1}-\mS^{(1)}_{\iC\iC}\right)^{-1}
    \mS^{(1)}_{\iC\iN} \mathbf{a}^{(1)}_\iN.
\label{eq:Scascade-bC}
\end{equation}
The power wave $\mathbf{b}^{(1)}_\iN$ exiting (1) via its free ports is then obtained by inserting Eq.\,(\ref{eq:Scascade-bC}) into Eq.\,(\ref{eq:Scascade-S1N}) using Eq.\,(\ref{eq:Scascade-eqa1}):
\begin{equation}
    \mathbf{b}^{(1)}_\iN = \mS^{(12)} \mathbf{a}^{(1)}_\iN,
\end{equation}
where we define the scattering matrix $\mathbf{S}^{(12)} \in \mathbb{C}^{n(\mathcal{N})\times n(\mathcal{N})}$ of the cascaded system:
\begin{equation}
    \mS^{(12)} = \mS^{(1)}_{\iN\iN} + \mS^{(1)}_{\iN\iC} 
    \left({\mS^{(2)}}^{-1} - \mS^{(1)}_{\iC\iC}\right)^{-1} 
    \mS^{(1)}_{\iC\iN}.
\label{eq:Scascade-final}
\end{equation}

In case $\mS^{(2)}$ is not invertible, Eq.\,(\ref{eq:Scascade-bC}) and Eq.\,(\ref{eq:Scascade-final}) can be rewritten using one of the following substitutions
\begin{subequations}
\begin{align}
    \left(\left({\mS^{(2)}}\right)^{-1} - \mS^{(1)}_{\iC\iC}\right)^{-1}
    & = \mS^{(2)} \left(\mI-\mS^{(1)}_{\iC\iC}\mS^{(2)}\right)^{-1} \\
    & = \left(\mI-\mS^{(2)}\mS^{(1)}_{\iC\iC}\right)^{-1} \mS^{(2)}.
\end{align}
\label{eq:substitutions}
\end{subequations}

The power waves $ \mathbf{a}^{(1)}_\iC = \mathbf{b}^{(2)}$ travelling across the connected ports from (2) to (1) can be evaluated directly based on Eq.\,(\ref{eq:Scascade-bC}) in terms of the power waves $\mathbf{a}^{(1)}_\iN$ impinging on the free ports of (1). Of course, the power waves $ \mathbf{a}^{(2)}=\mathbf{b}^{(1)}_\iC$ travelling across the connected ports from (1) to (2) are then simply obtained from $\mathbf{b}^{(2)}$ using Eq.\,(\ref{eq:Scascade-S2}).

It is hence furthermore possible to evaluate the power wave potentials at the connected ports and the power wave fluxes through the connected ports (see Appendix~\ref{Appendix_Definitions} for definitions).
Based on Eq.\,(\ref{def_eq_psi_phi}) and Eqs.~(\ref{eq:Scascade-eqa1},\ref{eq:Scascade-eqa1}), the power wave potentials $\boldsymbol{\psi}_\iC$ at the connected ports are given by
\begin{equation}
    \boldsymbol{\psi}_\iC = \mathbf{a}^{(1)}_\iC + \mathbf{b}^{(1)}_\iC = \mathbf{a}^{(2)} + \mathbf{b}^{(2)},
\end{equation}
and the power wave fluxes $\boldsymbol{\phi}_\iC^{(1)}$ flowing into $(1)$ at the connected ports are given by
\begin{equation}
    \boldsymbol{\phi}_\iC^{(1)} = -\boldsymbol{\phi}_\iC^{(2)} = \mathbf{a}^{(1)}_\iC - \mathbf{b}^{(1)}_\iC = \mathbf{b}^{(2)} - \mathbf{a}^{(2)},
\end{equation}
where $\boldsymbol{\phi}_\iC^{(2)}$ are the power wave fluxes flowing into $(2)$.
Using Eq.\,(\ref{eq:Scascade-bC}) and Eq.\,(\ref{eq:Scascade-S2}), we can directly link $\boldsymbol{\psi}_\iC$ and $\boldsymbol{\phi}_\iC^{(1)}$ to the input power wave $\mathbf{a}^{(1)}_\iN$:
\begin{equation}
    \boldsymbol{\psi}_\iC = \mathbf{\Psi}_{\iC\iN} \mathbf{a}^{(1)}_\iN, \quad
    \boldsymbol{\phi}_\iC^{(1)} = \mathbf{\Phi}_{\iC\iN} \mathbf{a}^{(1)}_\iN,
\end{equation}
where
\begin{equation}
    \mathbf{\Psi}_{\iC\iN}
    = \left( \left({\mS^{(2)}}\right)^{-1} + \mI \right)
    \left(\left({\mS^{(2)}}\right)^{-1} - \mS^{(1)}_{\iC\iC}\right)^{-1} \mS^{(1)}_{\iC\iN},
\label{eq:potential_matrix}
\end{equation}
\begin{equation}
    \mathbf{\Phi}_{\iC\iN}
    = \left( \left({\mS^{(2)}}\right)^{-1} - \mI \right)
    \left(\left({\mS^{(2)}}\right)^{-1} - \mS^{(1)}_{\iC\iC}\right)^{-1} \mS^{(1)}_{\iC\iN},
\label{eq:flux_matrix}
\end{equation}
or, using Eq.\,(\ref{eq:substitutions}) if $\mS^{(2)}$ is not invertible,
\begin{equation}
    \mathbf{\Psi}_{\iC\iN} = \left(\mI + \mS^{(2)} \right)\left(\mI - \mS^{(1)}_{\iC\iC}\mS^{(2)}\right)^{-1} \mS^{(1)}_{\iC\iN},
\end{equation}
\begin{equation}
    \mathbf{\Phi}_{\iC\iN} = \left(\mI - \mS^{(2)}\right) \left(\mI - \mS^{(1)}_{\iC\iC}\mS^{(2)}\right)^{-1} \mS^{(1)}_{\iC\iN}.
\end{equation}

\section{Analytic evaluation of a \\graph's scattering matrix}
\label{Appendix_graphs}

\newcommand{\va}{\ensuremath{\mathbf{a}}}
\newcommand{\vb}{\ensuremath{\mathbf{b}}}
\newcommand{\vphi}{\ensuremath{\boldsymbol{\phi}}}
\newcommand{\vpsi}{\ensuremath{\boldsymbol{\psi}}}
\newcommand{\mW}{\ensuremath{\mathbf{W}}}
\newcommand{\mM}{\ensuremath{\mathbf{M}}}
\newcommand{\mPhi}{\ensuremath{\mathbf{\Phi}}}
\newcommand{\mPsi}{\ensuremath{\mathbf{\Psi}}}
\newcommand{\cG}{\ensuremath{\mathcal{G}}}

This appendix summarizes the formalism used to compute the scattering matrix of a network of monomodal transmission lines, referred to as a graph in the present work. In this appendix, we follow the literature on scattering theory for the one-dimensional Schrödinger equation on a quantum graph~\cite{kostrykin_kirchhoffs_1999,texier_scattering_2001,kottos_quantum_2003, kuchment_quantum_2005}, focusing on power wave quantities. As stated in Sec.~\ref{subsec_val}, the same results can be derived in terms of the telegrapher's equations; the equivalence between the two approaches has been worked out explicitly in Ref.~\cite{hul_experimental_2004}.
Fundamentally, the derivation of the graph's scattering parameters relies on enforcing the continuity of the potentials $\psi$ and the conservation of the fluxes ${\phi}$ at the nodes \cite{kostrykin_kirchhoffs_1999}. 

\subsection{Scattering matrix of a graph}

\subsubsection{Power waves on graphs}

Solving the wave equation on the graph requires the expression of the potentials along the bonds (the wave solutions) as a function of those at the nodes. The potential $\psi_\alpha^\beta$ along the bond linking nodes $\alpha$ and $\beta$ and originating from the potential $\psi_\alpha$ of the node $\alpha$ must satisfy the following conditions\footnote{The full potential along the bond is the superposition of $\psi_\alpha^\beta$ and $\psi_\beta^\alpha$.}:
\begin{subequations}
\begin{align}
&\psi_\alpha^\beta\left(\alpha\right)=\psi_\alpha^\beta\left(x_{\alpha\beta}=0\right)=\psi_\alpha, \\&\psi_\alpha^\beta\left(\beta\right)=\psi_\alpha^\beta\left(x_{\alpha\beta}=l_{\alpha\beta}\right)=0,
\end{align}
\end{subequations}
where $l_{\alpha\beta}$ is the length of the bond between nodes $\alpha$ and $\beta$. For conciseness, we write $x$ instead of $x_{\alpha\beta}$ in the following. The fluxes can be decomposed into mono-directional components along the bonds. The one originating from $\psi_\alpha$ in the direction of $\beta$ is 
\begin{equation}
    \phi_\alpha^\beta(x) = -\frac{\jmath}{k} \nabla_x \psi_\alpha^\beta(x),
\end{equation}
where $k \in \mathbb{C}$ is the wavevector along the bonds of the graph and gradients are with respect to a position on the bond.

We define the nodal fluxes $\vphi$ as the sum of the  outgoing fluxes from a node $\alpha$ minus the incoming fluxes at $\alpha$ originating from every node $\beta$ bonded to $\alpha$:
\begin{equation}
    \phi_\alpha = \sum_{\beta \neq \alpha} c_{\alpha\beta}\Bigl(
    \phi_\alpha^\beta\left(\alpha\right) - \phi_\beta^\alpha\left(\alpha\right)
    \Bigr),
\end{equation}
where the connectivity $c_{\alpha\beta}=1$ if there is a bond linking the nodes $\alpha$ and $\beta$, and $c_{\alpha\beta}=0$ otherwise.

Casting the solution of Schrödinger's equation on a quantum wire as the mono-directional potential
\begin{equation}
\psi_\alpha^\beta\left(x\right) = \psi_\alpha \frac{\sin\left(k \left(l_{\alpha\beta}-x\right)\right)}{\sin\left(k l_{\alpha\beta}\right)}
\label{eq:schrodinger_solution}
\end{equation}
yields expressions for the outgoing flux from $\alpha$ and the incoming flux at $\beta$, both\footnote{$\phi_\alpha^\beta\left(\alpha\right)$ and $\phi_\alpha^\beta\left(\beta\right)$ are not necessarily equal due to the losses in the bond.} originating from $\psi_\alpha$:
\begin{subequations}
\begin{align}
    & \phi_\alpha^\beta\left(\alpha\right) = \jmath\psi_\alpha\cot{\left(kl_{\alpha\beta}\right)}, \\
    & \phi_\alpha^\beta\left(\beta\right) = \jmath\psi_\alpha\csc{\left(kl_{\alpha\beta}\right)}.
\end{align}
\label{eq:schrodinger_fluxes}
\end{subequations}

\subsubsection{Solving the wave equation}

Let us now consider an incoming power wave $\va$ impinging on the $N_\ell$ external nodes of a graph $\mathcal{G}$ comprising $N_n$ nodes in total. The matrix $\mPsi$ maps the incoming wave $\va$ to the potentials $\vpsi$ generated on every node:
\begin{equation}
    \vpsi = \mPsi \va.
\end{equation}
Moreover, we define the matrix $\mM$ such that it maps the potentials of the nodes to the nodal fluxes $\vphi$:
\begin{equation}
    \vphi = \mM \vpsi.
\label{eq:powerwave_M}
\end{equation}
Therefore, $\mM$ plays a role that is analogous to that of a nodal admittance matrix~\cite{monaco1974}, but regarding power waves potentials and fluxes rather than voltages and currents; it is hence constructed as follows:
\begin{equation}
    M_{\alpha\beta} = \left( \delta_{\alpha\beta} \sum_\gamma 
    c_{\alpha\gamma}\frac{\phi_\alpha^\gamma\left(\alpha\right)}{\psi_\alpha} 
    - c_{\alpha\beta}\frac{\phi_\beta^\alpha\left(\alpha\right)}{\psi_\beta}
    \right).
\end{equation}
Inserting the solution of Schrödinger's equation given in Eq.\,(\ref{eq:schrodinger_fluxes}) yields
\begin{equation}
    M_{\alpha\beta} = \jmath \delta_{\alpha\beta} \left[\sum_\gamma c_{\alpha\gamma} \cot \left( k l_{\alpha\gamma} \right) \right]- \jmath c_{\alpha\beta}\csc \left( k l_{\alpha\beta} \right).
\end{equation}

The potentials $\vpsi_\ell$ and nodal fluxes $\vphi_\ell$ at the externally connected nodes $\ell$ are the subsets of $\vphi$ and $\vpsi$ selected by the partial permutation matrix $\mW$ defined by
\begin{equation}
    \quad W_{\lambda\alpha} = \delta_{\lambda\alpha},
\end{equation}
where $\lambda$ spans the externally connected nodes $\ell$, $\alpha$ spans all the nodes $n$, and $\delta$ is the Kronecker delta. Specifically, 
\begin{equation}
    \vpsi_\ell = \mW \vpsi, \quad \vphi_\ell = \mW \vphi.
\label{eq:powerwave_W}
\end{equation}

The conditions relating the scattering matrix to the input potentials and nodal fluxes, similarly to Eq.\,(\ref{eq_Psi_I_S_a}), read
\begin{equation}
    \vpsi_\ell = \left(\mI+\mS\right)\mathbf{a}, \quad \vphi_\ell = \left(\mI-\mS\right)\mathbf{a}.
\label{eq:powerwave_S}
\end{equation}

The last condition to impose is the conservation of the fluxes on the exclusively internal nodes $n\setminus\ell$ of the graph. This is done by constraining the nodal fluxes of the internal nodes to zero, i.e.,
\begin{equation}
    \vphi_{n\setminus\ell} = \mathbf{0}.
\label{eq:powerwave_0}
\end{equation}

Combining Eqs.\,(\ref{eq:powerwave_M},\ref{eq:powerwave_W},\ref{eq:powerwave_S},\ref{eq:powerwave_0}), we obtain two coupled equations:
\begin{align}
    \mI - \mS & = \mathbf{W} \mPsi,     
\label{eq:graph_S_W_Psi} \\
    \mW^\mathrm{T} \left( \mI - \mS \right) & = \mM \mPsi.
\end{align}
These coupled equations can be solved for $\mPsi$, leveraging the fact that the pseudoinverse of $\mathbf{W}$ is equal to its transpose, yielding
\begin{equation}
    \mPsi = 2 \left(\mM + \mW^\mathrm{T}\mW \right)^{-1} \mW^\mathrm{T}.
\label{eq:graph_Psi}
\end{equation}
We obtain $\mS$ by reinserting Eq.\,(\ref{eq:graph_Psi}) into (\ref{eq:graph_S_W_Psi}):
\begin{equation}
    \mS = \mI - 2 \mW \left(\mM + \mW^\mathrm{T}\mW \right)^{-1} \mW^\mathrm{T}.
\end{equation}

\subsection{Recovery of fluxes and potentials at subgraph interfaces}

The bi-directional nodal fluxes $\phi_\alpha^\beta\left(\alpha\right)-\phi_\beta^\alpha\left(\alpha\right)$, at every node $\alpha$, along every bond of the graph $\mathcal{G}$, can be related to the input power wave $\va$ via the tensor $\mPhi$, defined as
\begin{equation}
    \Phi_{\alpha\beta\lambda} = c_{\alpha\beta} \Bigl(
    \frac{\phi_\alpha^\beta\left(\alpha\right)}{\psi_\alpha} \Psi_{\alpha\lambda}
    - \frac{\phi_\beta^\alpha\left(\alpha\right)}{\psi_\beta} \Psi_{\beta\lambda}
    \Bigr),
\end{equation}
and by inserting the solution of Schrödinger's equation from Eq.\,(\ref{eq:schrodinger_fluxes}):
\begin{equation}
    \Phi_{\alpha\beta\lambda} = \jmath c_{\alpha\beta} \Bigl(
    \cot\left(k l_{\alpha\beta}\right) \Psi_{\alpha\lambda}
    - \csc\left(k l_{\alpha\beta}\right) \Psi_{\beta\lambda}
    \Bigr).
\end{equation}

Consider a subgraph $\mathcal{S}$ comprising the nodes $s\subseteq\left(n\setminus\ell\right)$ of the full graph $\mathcal{G}$. Let us assume, for simplicity, that all bonds linking two nodes within $s$ are part of $\mathcal{S}$. It is possible to obtain, for example, the fluxes $\vphi^\mathcal{S}$ between $\mathcal{S}$ and $\mathcal{G}\setminus\mathcal{S}$ at the nodes $s$. One has to sum the bi-directional fluxes exchanged with every node in $n\setminus\ell$ at every node in $s$ by computing the matrix
\begin{equation}
    \mPhi^\mathcal{S} = \sum_{\beta\in\left(n\setminus\ell\right)}
    \Bigl( \Phi_{\alpha\beta\lambda} \mid \alpha \in s \Bigr),
\end{equation}
and obtain $\vphi^\mathcal{S}$ for a given input power wave $\va$ as
\begin{equation}
    \vphi^\mathcal{S} = \mPhi^\mathcal{S} \va.
\end{equation}
Meanwhile, having evaluated $\mathbf{\Psi}$ based on Eq.~(\ref{eq:graph_Psi}), the potentials $\vpsi^\mathcal{S}$ at the nodes connecting  $\mathcal{S}$ and $\mathcal{G}\setminus\mathcal{S}$ for a given input power wave $\va$ can be retrieved straightforwardly by selecting the corresponding entries of $\mathbf{\Psi}$. $\vphi^\mathcal{S}$ and $\vpsi^\mathcal{S}$ determine the ground truth required to verify our method from Sec.~\ref{sec:generalizedconnect_potentials}.

\section{Blockwise inversion of an ``off-diagonal-identity'' matrix}
\label{AppendixD}

Based on the block matrix inversion lemma, it can readily be verified that the following relations hold for the inverse of a generic $2 \times 2$ block matrix whose off-diagonal blocks are identity matrices:

\begin{equation}
\begin{split}
    \begin{bmatrix} \mathbf{A} & \mI \\ \mI & \mathbf{B} \end{bmatrix}^{-1}
    & = \begin{bmatrix} \mathbf{B} & -\mI \\ -\mI & \mathbf{A} \end{bmatrix}
    \begin{bmatrix} \left(\mathbf{AB-I}\right)^{-1} & \mO \\ \mO & \left(\mathbf{BA-I}\right)^{-1} \end{bmatrix} \\ 
    & = \begin{bmatrix}\left(\mathbf{BA-I}\right)^{-1} & \mO \\ \mO & \left(\mathbf{AB-I}\right)^{-1} \end{bmatrix}
    \begin{bmatrix} \mathbf{B} & -\mI \\ -\mI & \mathbf{A} \end{bmatrix}.
\end{split}
\label{eq:block_inv_I}
\end{equation}

\bibliographystyle{IEEEtran}

\providecommand{\noopsort}[1]{}\providecommand{\singleletter}[1]{#1}%

\end{document}